\documentclass[twocolumn,aps,prd,unsortedaddress,%
superscriptaddress,a4paper,nofootinbib,balancelastpage,preprintnumbers,floatfix,longbibliography]{revtex4-1}
\usepackage{natbib}
\usepackage[utf8]{inputenc}
\usepackage{latexsym,amsbsy,amsmath}
\usepackage{mathtools}
\usepackage{graphicx}
\usepackage{xcolor}
\usepackage{longtable}
\usepackage{isotope}
\usepackage{microtype}
\usepackage[bitstream-charter]{mathdesign}
\renewcommand{\vec}[1]{\boldsymbol{#1}}

\def\ll#1#2{\tilde{\lambda}_{#1}.\tilde{\lambda}_{#2}}

\usepackage{palatino}

\usepackage{float}
\makeatletter
\newcommand*{\rom}[1]{\expandafter\@slowromancap\romannumeral #1@}
\makeatother
\begin{document}
\title{Few-body quark dynamics for  doubly-heavy baryons and tetraquarks}
\author{Jean-Marc~Richard}
\email{j-m.richard@ipnl.in2p3.fr}
\affiliation{Universit\'e de Lyon, Institut de Physique Nucl\'eaire de Lyon,
IN2P3-CNRS--UCBL,\\
4 rue Enrico Fermi, 69622  Villeurbanne, France}
\author{Alfredo~Valcarce}
\email{valcarce@usal.es}
\affiliation{Departamento de F{\'\i}sica Fundamental and IUFFyM,
Universidad de Salamanca, 37008 Salamanca, Spain}
\date{\emph{Version of }\today}
\author{Javier~Vijande}
\email{javier.vijande@uv.es}
\affiliation{Unidad Mixta de Investigación en Radiofísica e Instrumentación Nuclear en Medicina (IRIMED), Instituto de Investigación Sanitaria La Fe (IIS-La Fe)-Universitat de Valencia (UV) and IFIC (UV-CSIC), Valencia, Spain}
\begin{abstract}
\noindent
We discuss the adequate treatment of the 3- and 4-body dynamics for the quark model picture of double-charm baryons and tetraquarks. We stress that the variational and Born-Oppenheimer approximations give energies very close to the exact ones, while the diquark approximation might be rather misleading. The Hall-Post inequalities also provide very useful lower bounds that exclude the possibility of stable tetraquarks for some mass ratios and some color wave functions. 
\end{abstract}
\maketitle
\section{Introduction}
\label{se:intro}
There is rich literature on multiquarks, and many reviews,  including \cite{Briceno:2015rlt,*Lebed:2016hpi,*Chen:2016qju,*Ali:2017jda,*Esposito:2016noz,*Richard:2016eis}. The recent contributions are stimulated by the discovery of a double-charm baryon \cite{Aaij:2017ueg}, which is interesting by itself and also triggers speculations about exotic double-charm mesons $QQ\bar q\bar q$. For years, the sector of flavor-exotic tetraquarks has been somewhat forgotten, and even omitted from some reviews on exotic hadrons,
as much attention was paid to  hidden-flavor states $Q\bar Q q\bar q$. However the flavor-exotic multiquarks have been investigated already some decades ago~\cite{Ader:1981db} and has motivated an abundant literature~\cite{Ballot:1983iv,*Zouzou:1986qh,*Heller:1985cb,*Carlson:1988hh,*Heller:1986bt,*Brink:1994ic,*Brink:1998as,*Vijande:2003ki,*Janc:2004qn,*Vijande:2007rf,*Vijande:2007ix,*Carames:2011zz,*Hyodo:2012pm,*Mehen:2017nrh,*Yasui:2013tsa,*Czarnecki:2017vco}  that has been unfortunately ignored in some recent papers. 

The underlying dynamics is not exactly the same in all papers cited in \cite{Ballot:1983iv}. Some authors consider a purely linear interaction, either pairwise or inspired by the string model, and some others include a Coulomb-like interaction and spin-dependent terms. Sometimes, the wave function contains a single color configuration, while in other papers the role of color mixing is analyzed.

In the present note, we stress that a careful treatment of the few-body problem is required before drawing any conclusion about the existence of stable states in a particular model. We, indeed, observe a dramatic spread of strategies: some authors use the full machinery of a variational method based on correlated Gaussians or hyperspherical expansion, and others use a crude trial wave function or a cluster approximation. We shall review critically the different strategies that can be found in the literature. 

Not surprisingly, the main difficulties are encountered when a multiquark state is found near its lowest dissociation threshold. The question of whether or not there is a bound state requires a lot of care.  In particular, one should account for the mixing of color configurations~\cite{Richard:2017vry,Vijande:2009zs}.

We apologize for the somewhat technical character of this survey. However, we find it necessary to clarify the somewhat contradictory results in the literature. In particular, some authors who use similar ingredients obtain either stability or instability for the all-heavy configuration $QQ\bar Q\bar Q$, and in our opinion, this is due to an erroneous handling of the four-body problem in some papers.  

This paper is organized as follows. In Sec.~\ref{se:var}, we briefly discuss the variational approximation, with several variants, including the hyperspherical expansion. In Sec.~\ref{se:diq}, we discuss the diquark approximation, that is widely used. In Sec.~\ref{se:BO}, we discuss the Born-Oppenheimer method. In Sec.~\ref{se:EQ}, we comment about the approximate relation between meson, baryon and tetraquark energy. In Sec.~\ref{se:HP} a reminder is given  about the Hall-Post inequalities, and some new applications are derived for tetraquarks within potential models. The importance of color mixing is illustrated in Sec.~\ref{se:col-mix}. 
The role of the spin-dependent part of the potential is stressed in Sec.~\ref{se:spin}.  Some conclusions are proposed in~Sec.~\ref{se:concl}. 
\section{Variational methods}\label{se:var}
\subsection{General considerations}
Variational methods have been applied from the beginning of  quantum mechanics, as they were already used in other fields of physics involving similar equations. A well-known example is the Helium atom, for which the unperturbed  wave function $\Psi_0=\exp(-2\,r_1-2\,r_2)$, in an obvious notation, is already a good trial function, and can be improved, without much further computation, in the form $\Psi(\alpha)=\exp(-\alpha\,r_1-\alpha\,r_2)$, where $\alpha$ is empirically adjusted, and is interpreted as the effective charge seen by each electron. See, e.g., \cite{1957qmot.book.....B,*2010AmJPh..78...86H}.

However, the stability of $\alpha e^- e^-$ is obvious as once the first electron is bound, there is enough attraction left to attach the second one. More delicate is the case of H$^-(p e^- e^-)$, for which the above trial function does not achieve binding, nor any factorized $f(r_1)\,f(r_2)$.  As shown by Hylleraas, and independently by Chandrasekhar  (see refs. in \cite{1957qmot.book.....B}), achieving binding requires either some asymmetry and restoration of symmetry, as $\exp(-\alpha\,r_1-\beta\,r_2)+ [\alpha\leftrightarrow\beta]$, or some explicit anticorrelation, such as $\exp(-\alpha\,r_1-\alpha\,r_2-\gamma\,r_{12})$, or, of course, a combination of both. 

Similarly, the energy and structure of a baryon is easily calculated in any quark model, as the wave function is rather compact. But for a tetraquark $q_1q_2 \bar q_3\bar q_4$ at the edge of binding, the wave function contains antibaryon-like components with $q_1q_2$ clustered, meson-meson components such as $q_1\bar q_3-q_2\bar q_4$, and perhaps some diquark-antidiquark contributions. Thus a simplistic variational function cannot account for these three aspects. 

For illustration, we shall use some toy models with increasing complexity. In the simplest version, the color wavefunction is frozen as $\bar 33$ in the $(QQ)(\bar q\bar q)$ basis, and the potential is purely chromoelectric. It reads
\begin{multline}\label{eq:H4}
 H_{33}=\frac{\vec p_1^2+\vec p_2^2}{2\,M}+\frac{\vec p_3^2+\vec p_4^2}{2\,m}\\{}+\frac{v_{12}+v_{34}}{2}+\frac{v_{13}+v_{14}+v_{23}+v_{24}}{4}~,
\end{multline}
where $v_{ij}=v(r_{ij})$, with $v(r)$ being either $r$ or $r^{0.1}$ or $\lambda\,r-\kappa/r$ for illustration. The analog with color $6\bar 6$ reads
\begin{multline}\label{eq:H4-6}
 H_{66}=\frac{\vec p_1^2+\vec p_2^2}{2\,M}+\frac{\vec p_3^2+\vec p_4^2}{2\,m}\\{}-\frac{v_{12}+v_{34}}{4}+\frac58\,(v_{13}+v_{14}+v_{23}+v_{24})~.
\end{multline}
If color mixing is accounted for, then one gets a coupled-channel problem
\begin{equation}\label{eq:H4-36}
H=\begin{pmatrix}
   H_{33}&H_{36}\\
   H_{36}&H_{66}
  \end{pmatrix}~,\ H_{36}=\frac{3(v_{14}+v_{23}-v_{13}-v_{24})}{4\,\sqrt2}~. 
\end{equation}

It can be checked, that a simple one-Gaussian wave function $\exp(-a\,\vec x^2-b\,\vec y^2-c\,\vec z^2)$, where
\begin{equation}\label{eq:Jacobi}
\vec x=\vec r_2-\vec r_1~,\quad
\vec y=\vec r_4-\vec r_3~,\quad
\vec z=\frac{\vec r_3+\vec r_4-\vec r_1-\vec r_2}{\sqrt2}~, 
\end{equation}
is a set of Jacobi variables, describes rather well the ground state of the  single-channel Hamiltonians $H_{33}$  or $H_{66}$. As reviewed in \cite{Richard:2017vry}, for $M=m$, the $6\bar 6$ is lower than the $\bar 33$ one. For $M/m\gg1$, the $\bar 33$ channel benefits from the $QQ$ attraction, and becomes more favorable. However, by itself, it requires a large value of $M/m$ to achieve stability below the $2\,Q\bar q$ threshold. The critical value $(M/m)_c$ depends on the shape of the potential, for instance $(M/m)_c\gtrsim 40$ for a linear interaction, $(M/m)_c\sim 15$ for a soft potential $r^{0.1}$ and $(M/m)_c\sim 7$ for an attractive Coulomb interaction. 

\begin{figure}[!ht]
 \centering
 \includegraphics[width=.7\columnwidth]{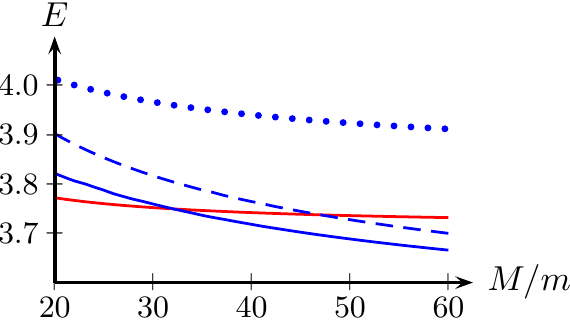}
 \caption{Comparison of different approximations for a tetraquark bound by a linear potential with frozen color wavefunction, or with color mixing. Dotted blue line: pure $6\bar 6$, dashed blue line: pure $\bar33$, solid blue line: with color mixing, red line: threshold. The units are such that $m=1$ and $v_{ij}=r_{ij}$.}
 \label{fig:BilanL}
\end{figure}

This critical value $(M/m)_c$ is significantly lowered if one refines the wave function and introduces color mixing, i.e., uses $H$ instead of $H_{33}$ alone. Due to the different symmetry patterns of the color $\bar 33$ and $6\bar 6$ states, the mixing requires an antisymmetric (under $1\leftrightarrow2$ or $3\leftrightarrow4$) wavefunction in one of the channels. The minimal wave function is thus
\begin{multline}\label{eq:psi-sym-asym}
 \Psi\propto \exp\bigl[-a_u\,r_{12}^2 -b_u\,r_{34}^2-c_u\,(r_{13}^2+r_{24}^2)\\
 {}-c'_u\,(r_{14}^2+r_{23}^2)\bigr] \pm (c_u\leftarrow c'_u)~,
\end{multline}
where $u$ stands for $\bar33$ or $6\bar6$, and $c_u\neq c_u'$ in the antisymmetric channel ($6\bar6$ in practice).
The effect of color mixing is illustrated in Fig.~\ref{fig:BilanL} (using a simple variational method, so that the actual energy might be slightly lower).
It is seen that the critical value for binding is reduced to  $(M/m)_c\sim 32$ by color mixing.\footnote{In Fig.~\ref{fig:BilanL} and similar figures, the energies above the threshold are an artifact of any variational calculation based on normalizable wavefunctions. The proper treatment of the continuum requires dedicated techniques.}

\begin{figure}[!hb]
 \centering
 \includegraphics[width=.7\columnwidth]{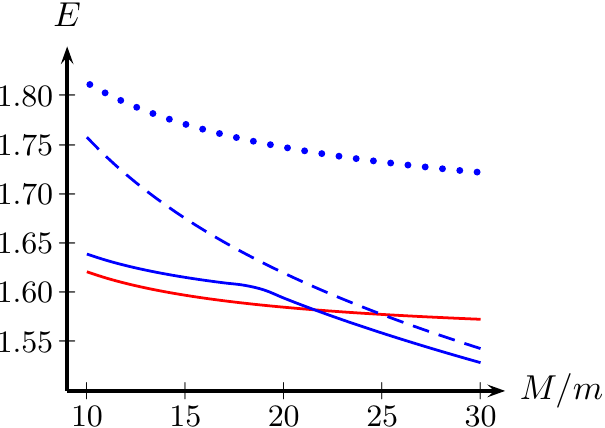}
 \caption{Same as Fig.~\ref{fig:BilanL} for a Coulomb-plus-linear interaction $v(r)=-0.4/r+0.2\,r$, where $r$ is in GeV$^{-1}$. The energy $E$ is in GeV.}
 \label{fig:BilanLC}
\end{figure}

The effect of an explicit Coulomb part in the spin-independent potential is seen in Fig.~\ref{fig:BilanLC}. The potential is chosen as $v(r)=-\kappa/r+\sigma\,r$ with $\kappa=0.4$ and $\sigma=0.2\,\mathrm{GeV}^2$. The light mass is taken as $m=0.3\,$GeV.
%
One remarks that the effect of color mixing is less dramatic; the explicit inclusion of a Coulomb term decreases the critical value $(M/m)_c$ significantly, here about 18 instead of about 28 (with a simple Gaussian expansion). 

We shall return in Sec.~\ref{se:col-mix} to the problem of color mixing, with more realistic models that include a spin-spin component. 
%
%
\subsection{Correlated Gaussian expansion}
A more efficient wave function is 
\begin{equation}
 \label{eq:var1}
 \Psi=\sum_i \gamma_i \exp(-a_{11,i}\,\vec x^2- 2\,a_{12,i}\,\vec x.\vec y- \cdots - a_{33,i}\vec z^2)~,
\end{equation}
which describes an overall scalar with the possibility of internal orbital excitations. The quadratic form  $a_{11,i}\,\vec x^2+ 2\,a_{12,i}\,\vec x.\vec y+\cdots$ is positive-definite, and is sometimes rewritten as
$\sum_{j<k} b_{jk,i}\,r_{jk}^2$  
with all $b_{jk,i}$ positive. None of the Gaussians fulfill the requirements of permutation symmetry, but $\Psi$ does, after optimization of the parameters. 

A variant of \eqref{eq:var1} consists of using only diagonal Gaussians associated to the coordinates $\vec x$, $\vec y$ and $\vec z$, but to add  diagonal terms in other sets of Jacobi coordinates, say
\begin{multline}
 \label{eq:var2}
 \Psi=\sum_i d_i \exp(-a_i\,\vec x^2-b_i\,\vec y^2- c_i\,\vec z^2)\\
 {}+\sum_i d'_i \exp(-a'_i\,\vec x^2-b'_i\,\vec y^2- c'_i\,\vec z^2)+\cdots~,
\end{multline}
where, for instance, 
\begin{equation}\label{eq:Jacobip}
\begin{gathered}
\vec x'=\vec r_3-\vec r_1~,\qquad
\vec y'=\vec r_4-\vec r_2~,\\
\vec z'\propto M\,\vec r_2+m\,\vec r_4-M\,\vec r_1-m\,\vec r_3~, 
\end{gathered}
\end{equation}
corresponding to different cluster decompositions \cite{Hiyama:2003cu}. In this case, the spin-isospin-color algebra is slightly more delicate.

Other variants deal with the numerical determination of the parameters. For a given set of range parameters the weights $\gamma_i$ in \eqref{eq:var1} or $d_i, d'_i,\ldots$ in \eqref{eq:var2} and the energy are given by a generalized eigenvalue equation. The range parameters themselves are searched for by stochastic methods \cite{2013RvMP...85..693M} or as belonging to a geometric series \cite{Hiyama:2003cu}. In both cases, the method is now well functioning. 
\subsection{Hyperspherical expansion}
By properly rescaling the Jacobi coordinates $\vec x$, $\vec y$, \dots, the Hamiltonian describing the relative motion of the quarks can be written as
\begin{equation}\label{eq:HH1}
H=\frac{1}{\mu}(\vec p_x^2+\vec p_y^2+\cdots)+V(\vec x,\vec y,\ldots)~, 
\end{equation}
which can be read as a  Schrödinger equation for a single particle of mass $\mu/2$ in a world of spatial dimension $3\,(n-1)$, where $n=3$ for baryons, $n=4$ for tetraquarks, etc. 
In general, the potential $V(\vec x,\vec y,\ldots)$ is not central, so the partial wave expansion 
\begin{equation}\label{eq:HH2}
 \Psi=\sum_{[L]} R_{[L]}(r) \mathcal{Y}_{[L]}(\Omega)~,
\end{equation}
results into an infinite set of coupled  equations for the radial functions $R_{[L]}(r)$ or their reduced form $r^{5/2}\,R_{[L]}(r)$. But if one solves with an increasing number of equations, the convergence is rather fast. Here, $r=(\vec x^2+\vec y^2+\cdots)^{1/2}$ is the hyperradius, $\Omega$ a set of $3\,n-4$ angles, and $[L]$ denotes the ``grand'' angular momentum $L$  and its associated magnetic numbers labeling the generalized spherical harmonics $\mathcal{Y}$. 

The convergence is illustrated in Table~\rom{9} and Fig.~2 of Ref.~\cite{Vijande:2009kj}. 
\section{Diquark approximation}\label{se:diq}
The motivations for diquarks cover much more than hadron spectroscopy. See, e.g., \cite{Anselmino:1992vg} for a survey and references to pioneering articles which are sometimes ignored in the recent literature. A few decades ago, the main concern in baryon spectroscopy, was the problem of missing resonances, predicted by the quark model and not observed.  Many states of the symmetric quark model disappear if baryons are constructed out of a frozen diquark and a quark. However, the missing resonances, in which of the degrees of freedom $\vec x=\vec r_2-\vec r_1$ and $\vec y\propto \vec r_3 -(\vec r_1+\vec r_2)/2$ are both excited, are not very much coupled to the typical investigation channels $\pi N$ or $\gamma N$ which privilege states with one pair of quarks shared with the target nucleon $N$. In recent photoproduction experiments with improved statistics, some of the missing states have been identified, which cannot be accommodated as made of a ground-state diquark and a third quark \cite{Klempt:2017lwq}. So one of the grounds of the diquark model is somewhat weakened. 

The diquark model is regularly revisited, to accommodate firmly established exotics such a the $X(3872)$, or candidates awaiting confirmation~\cite{Briceno:2015rlt}. Unfortunately, some unwanted multiquarks are also predicted in this approach, though this is not always explicitly stated or even realized.  The issue of multiquarks within the diquark model was raised many years ago by Fredriksson and Jandel \cite{Fredriksson:1981mh}\footnote{Some technical details of that paper might be revised, but the main concern remains.}, and is sometimes  rediscovered, without any reference to the 1981 paper.  The paradox is perhaps that the diquark model, that produces fewer baryon states, produces too many multiquarks!

There are  many variants of the so-called diquark model. An extreme point of view is that diquarks are almost-elementary objects, with their specific interaction with quarks and between them. A whole baryon phenomenology can be built starting from well-defined assumptions about the diquark constituent masses and the potential linking a quark to a  diquark. Then a diquark-diquark interaction has to be introduced as a new ingredient for the multiquark sector.  

Another extreme is to view diquarks as a type of ``Voodoo few-body\footnote{Jaffe \cite{Jaffee:1986zz} reported  that Bjorken used the words ''Voodoo QCD'' to denote several useful models of strong-interaction physics, such a vector meson-dominance, and also some less convincing recipes. A correspondence with R.L.~Jaffe is gratefully acknowledged.}''. In this empirical approach to few-body physics, to estimate the energy and wave function of, say, $(a_1 a_2 a_3)$, with masses $m_i$ and interaction $v_{ij}(r)$, one first solves for $(a_1 a_2)$ with $v_{12}$ alone, with energy $\eta_{12}$, then estimates the bound state of a point-like $(a_1a_2)$ of mass $m_1+m_2$, or perhaps $m_1+m_2+\epsilon_2$ in some variants, located at $\vec R_{12}$, interacting with $a_3$ through the potential $v_{13}(\vec r_3-\vec R_{12})+v_{23}(\vec r_3-\vec R_{12})$, resulting in binding energy $\eta_{12,3}$, and the whole  energy is given by $\eta_{12}+\eta_{12,3}$. For a 4-body system, the $(a_1a_2)$ and $(a_3a_4)$ systems are estimated first, and then a third two-body equation is solved for $(a_1a_2)$ interacting with $(a_3a_4)$ via a potential $\sum' v_{ij}(\vec R_{12}-\vec R_{34})$, where $\Sigma'$ denotes $i=1,2$ and $j=3,4$ throughout this paper, in particular in Sec.~\ref{se:HP}.

This strategy is of course fully justified for the deuterium atom considered as a $pne^-$ system, as the inter-nuclear motion is not significantly modified by the electron. On the other hand, this approach ruins some subtle collective binding, for instance, that of Borromean states~\cite{2006FBS....38...57F}.
Also, one cannot see either how H$^-(pe^- e^-)$ could become bound in this approach, or  the hydrogen molecule be described as a ''diproton'' linked to a ``dielectron''!
In other cases, the method  just underestimates the binding: for a $\alpha e^- e^-$ atom with a static nucleus, the first electron would  get an energy $-2$ in natural units, and the second, only an energy $-0.5$, as it would endorse a full screening, while the exact energy is about $-2.90$. For the quark model, the effect is opposite, and, as seen below, the \textsl{ad-hoc} clustering lowers significantly the energy. 
\subsection{The diquark model for double-charm baryons}
\begin{figure}[ht]
 \centering
 \includegraphics[width=.3\textwidth]{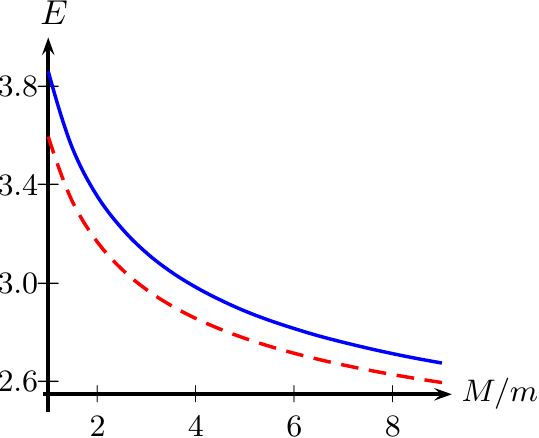}
 \caption{Comparison of the exact energy (solid blue line) and diquark approximation (red dotted line) for a baryon $(QQq)$ with masses $M$ and $m=1$, and a purely linear interaction, as a function of the mass ratio.  The units are such that $=1$ and the potential is  $\sum r_{ij}/2$. }
 \label{fig:QQq}
\end{figure}

In the case of double-charm baryons $QQq$ there is obviously a $QQ$ clustering which makes it tempting to use a two-step approach:  first a $(QQ)$ diquark and then a $(QQ)q$ quasi-meson, as the diquark has the same color $\bar 3$ as an antiquark. In Fig.~\ref{fig:QQq}, we compare the exact energy of $QQq$ bound by a linear interaction $\sum r_{ij}/2$ and the diquark approximation, as a function of the heavy-to-light quark mass ratio $M/m$. 

%
There is a clear overbinding. The situation does not improve too much as $M/m$ increases:  the diquark becomes more compact, but simultaneously, the total energy is more and more dominated by the heavy sector, so any systematic error in the $QQ$ effective interaction is more visible. The problem, as already stressed in~\cite{Richard:2017vry}, is that the light quark induces some interaction between the two heavy quarks. In the case of an harmonic confinement, $V=\sum r_{ij}^2/2$, the potential splits exactly into $V=3\,(\vec x^2+\vec y^2)/4$ if the second Jacobi variable is normalized as 
$\vec y=(2\,\vec r_3-\vec r_1-\vec r_2)/2$. 
The naive diquark approximation consists of replacing $3\,x^2/4$ by $x^2/2$, so that the contribution of the heavy quarks to the energy is reduced by a factor $(3/2)^{1/2}$. Similarly, for a linear interaction, the light quark potential, averaged over a sphere surrounding the diquark, will induce a positive contribution which is either $\propto x^2/y$ or $\propto y^2/x$, depending on the radius, and is omitted in the naive diquark model.  
\subsection{The diquark model for doubly-heavy tetraquarks}
\begin{figure}[ht!]
 \centering
 \includegraphics[width=.35\textwidth]{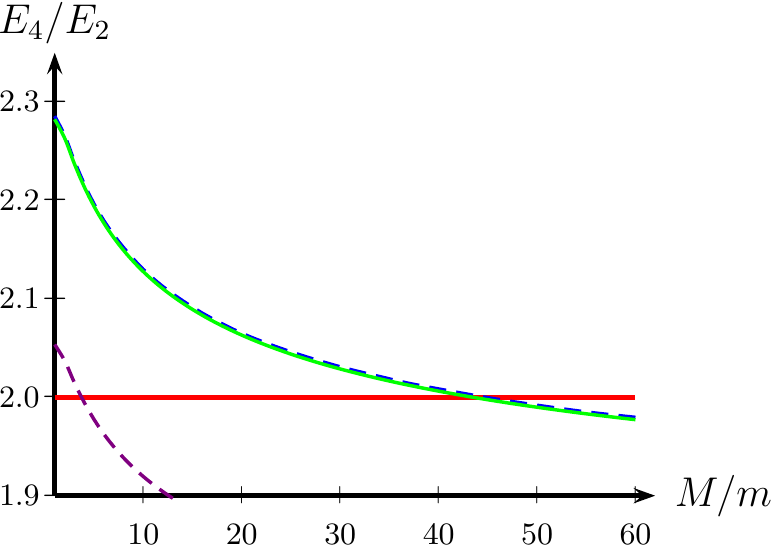}
 \caption{Comparison of the variational upper bound (green curve) and Hall-Post lower bound (dotted blue curve), hardly distinguishable from the variational estimate at this scale, for the tetraquark Hamiltonian \eqref{eq:H4} with a linear interaction. Also shown is the naive diquark-antidiquark approximation (dashed violet curve).}
 \label{fig:HP1}
\end{figure}

\begin{figure}[ht!]
 \centering
 \includegraphics[width=.35\textwidth]{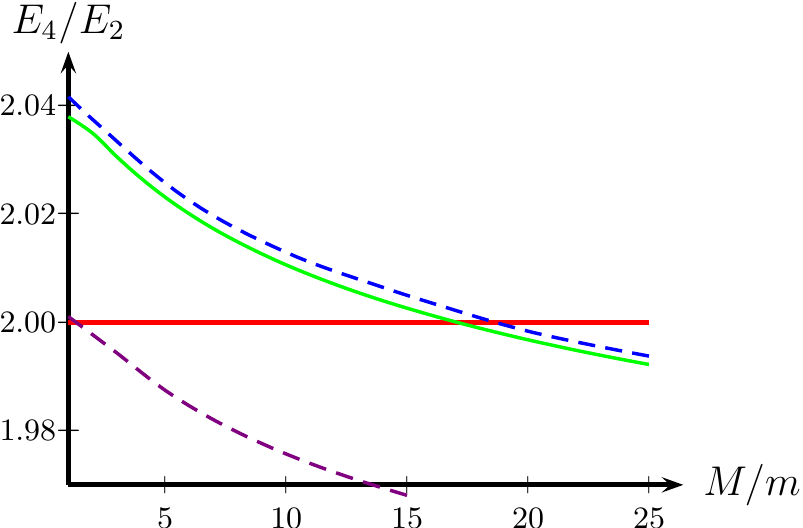}
 \caption{Same as Fig.~\ref{fig:HP1} with a potential $r_{ij}^{0.1}$ instead $r_{ij}$.}
 \label{fig:HP01}
\end{figure}

\begin{figure}[ht!]
 \centering
 \includegraphics[width=.35\textwidth]{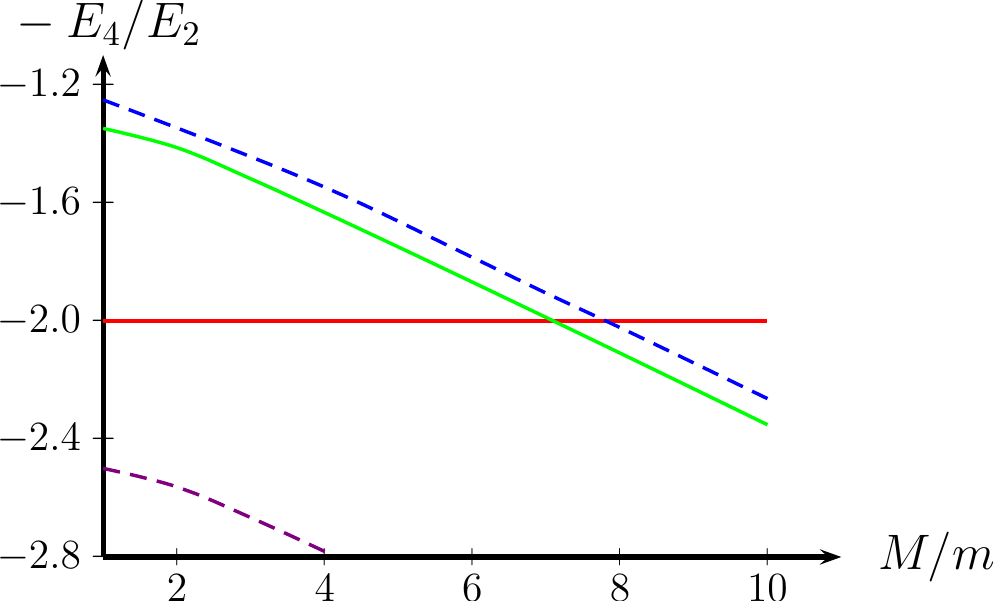}
 \caption{Same as Fig.~\ref{fig:HP1} with a potential $-r_{ij}^{-1}$ instead $r_{ij}$. The threshold is fixed at $-2$.}
 \label{fig:HPC}
\end{figure}

The exercise can be repeated for the $QQ\bar q\bar q$ states. For simplicity, we consider only the case of a frozen $\bar 33$ color wave function, i.e., the Hamiltonian \eqref{eq:H4}. Color mixing has to be introduced to have the proper threshold in the model, and it has been seen in explicit calculations that the mixing with meson-meson configurations is crucial for states at the edge of stability. Nevertheless the comparison of various approximations is instructive for the toy model \eqref{eq:H4}. In Fig.~\ref{fig:HP1}, we compare the exact solution of \eqref{eq:H4} with the approximation consisting of first computing the $QQ$ diquark with $r_{12}/2$ alone and $qq$ with $r_{34}$ alone, and then $(QQ)(\bar q\bar q)$ as a meson with a potential $r_{12,34}$ and constituent masses $2\,M$ and $2\,m$. The comparison is also made for a soft interaction $r^{0.1}$ in Fig.~\ref{fig:HP01} and a pure Coulomb interaction in Fig.~\ref{fig:HPC}. 
%
%
\section{Born-Oppenheimer method}\label{se:BO}
\subsection{General considerations}
The Born-Oppenheimer method is implicit in any quark model. The quarkonium potential, for instance, is the minimal energy of the gluon field for a given separation of the quark and the antiquark. Explicit reference to Born-Oppenheimer was made, e.g., in the context of the bag model \cite{Hasenfratz:1977dt}. Then it was speculated that some exotic mesons are just quarkonia evolving in a color field with gluonic or light-quark pairs excitations, see, e.g.,  \cite{Hasenfratz:1980jv,*Braaten:2014qka}.

For a given interquark potential, there is also a  Born-Oppenheimer approximation (BOA) for the solution of the Schrödinger equation governing double-charm baryons or double-charm tetraquarks, in analogy with the treatment of H$_2{}^+$ and H$_2$ in atomic physics, and it works very well, even for moderate values of the quark mass ratio $M/m$. 

Actually, in the most naive version of BOA, the heavy quarks are frozen, and the energy of the light quark(s), supplemented by the direct $QQ$ interaction, provides an effective potential that is independent of $M$. For finite $M$, the most significant correction comes from the recoil of the heavy quarks. This correction disappears if one applies BOA on the intrinsic Hamiltonian, free of center-of-mass motion. More precisely, in the case of baryons, 
let us consider
\begin{equation}\label{eq:BOA1}
 H_3=\frac{\vec p_x^2}{M}+\frac{\vec p_y^2}{\mu}+V(\vec x,\vec y)~,
\end{equation}
and search the solution as 
\begin{equation}\label{eq:BOA2}
 \Psi=\varphi(\vec x)\,\psi(\vec x,\vec y)
\end{equation}
where $\psi(\vec x,\vec y)$ is the solution of the one-body equation
\begin{equation}\label{eq:BOA3}
 -\frac{\Delta_{y}\psi(\vec x,\vec y)}{\mu}+V(\vec x,\vec y)\psi(\vec x,\vec y)=\epsilon(x)\,\psi(\vec x,\vec y)~.
\end{equation}
The BOA consists of neglecting in the kinetic energy operator the variations of $\psi$ as a function of $\vec x$, and to deduce the first levels from 
\begin{equation}\label{eq:BOA4}
 -\frac{\Delta_{x}\varphi(\vec x)}{\mu}+\epsilon(x) \, \varphi(x)=E\,\varphi(\vec x)~.
\end{equation}
The ground state energy is underestimated (i.e., binding overestimated), as the last two terms of \eqref{eq:BOA1} are replaced by their minimum\footnote{These considerations can be extended to the excited states: the sum of $n$ first levels is underestimated by BOA.}. Note that if the wavefunction \eqref{eq:BOA2} is used as a trial function, one gets an upper bound for the ground-state, sometimes named ''variational Born-Oppenheimer''. 
\subsection{Born-Oppenheimer for baryons}
The validity of BOA for $QQq$ baryons was shown in~\cite{Fleck:1989mb}. The check below  is just for completeness. The light-quark energy $\epsilon(x)$ can be calculated  by ordinary partial-wave expansion, which leads to coupled radial equations. One can also use a variational method, namely
\begin{equation}\label{eq:BOB1}
 \psi(\vec x,\vec y)=\sum_i \gamma_i\left[ \exp(-a_i\,\vec y^2-b_i\,\vec y.\vec s_i)+\vec s_i\leftrightarrow -\vec s_i)\right]~,
\end{equation}
where $\vec s_i\parallel \vec x$. The matrix elements of the normalization, kinetic energy and potential energy are given in a recent compilation \cite{Fedorov:2017bcq}.
The light-quark energy $V_q=\epsilon(x)-x/2$ is shown in Fig.~\ref{fig:BO34}, in the case of a linear potential.  For $x=0$, the result is analytic. 
%
%
%
\subsection{Born-Oppenheimer for tetraquarks}
Here, once more, we use the toy Hamiltonian \eqref{eq:H4}. It corresponds to a frozen $\bar 33$ color wavefunction. The effective potential is estimated using a trial wave function that generalizes \eqref{eq:BOB1} as to include two Jacobi coordinates, $\vec y$ and $\vec z$ in the light sector. For $x=0$, the light quark energy $V_q=\epsilon(x)-x/2$ coincides with the energy of a singly-heavy baryon $Q'qq$ with a flavored quark of mass $M'=2\,M$. This provides a check of the numerics. We shall come back to this point in Sec.~\ref{se:EQ}. 
The light-quark energy is shown in Fig.~\ref{fig:BO34}. 
%
%
\section{Relating mesons, baryons and tetraquarks}\label{se:EQ}
In a recent paper, Eichten and Quigg \cite{Eichten:2017ffp} use the heavy-quark symmetry to relate meson, baryon and tetraquark energies. In a simplified version without spin effects, it reads
\begin{equation}
 \label{eq:EQ1}
 QQ\bar q\bar q=QQq + Qqq-Q\bar q~,
\end{equation}
where the configuration stands for the ground-state energy. For fixed $m$ and $M\to \infty$, the identity is exact. For finite $M$, there is some departure. For instance with a purely linear model, in units such that  $v(r)=r$ for mesons, $\sum_{i<j} r_{ij}/2$ for baryons, $m=1$ and $M=5$ in the Hamiltonian \eqref{eq:H4} with frozen $\bar33$ color for tetraquarks, one gets 4.331 for he l.h.s.\  and 4.357 for the r.h.s.\ of \eqref{eq:EQ1}.
If one treats the tetraquark $QQ\bar q\bar q$ and the doubly-heavy baryon $QQq$ in the Born approximation, one can compare the two effective potentials as a function of the $QQ$ separation $x$, the baryon one being shifted by $ Qqq-Q\bar q$ which is independent of $x$. Without recoil correction, the two potentials are identical at $x=0$. For finite $M$, there is slight difference, as the single $q$ recoils against either $M$ or $2\,M$, and similarly $qq$ recoils against one or two heavy quarks.

\begin{figure}[ht]
 \centering
\includegraphics[width=.35\textwidth]{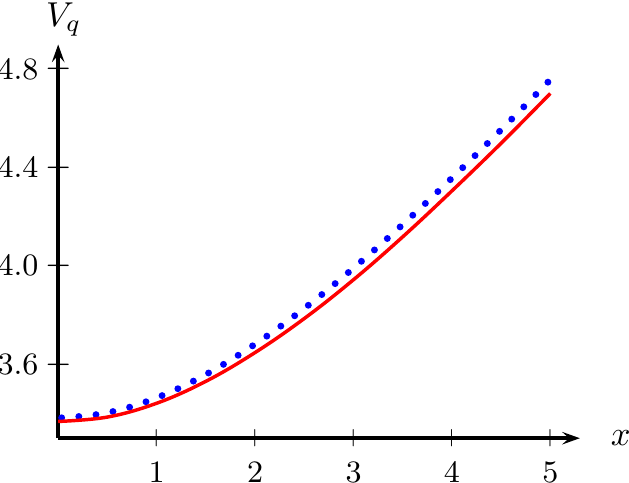}
 \caption{Comparison of the light quark energies for $QQ\bar q\bar q$ (solid red line) and $QQq$ (dotted blue line) as a function of the $QQ$ separation~$x$. The second curve is shifted by the difference of energies $Qqq-Q\bar q$. The units are such $m=1$, $M=5$ and $v_{ij}=r_{ij}$.}
 \label{fig:BO34}
\end{figure}

The comparison is shown in Fig.~\ref{fig:BO34}. Clearly the two effective potentials are very similar, and thus give almost identical energies, up to an additive constant that corresponds to the last two terms in \eqref{eq:EQ1}.
\section{Hall-Post inequalities}\label{se:HP}
\subsection{A brief reminder}
The Hall-Post inequalities have been derived in the 50s to relate the binding energies of light nuclei with different number of nucleons \cite{1967PPS....90..381H}. They have been re-discovered in the course of studies on the stability of matter  \cite{1966JMP.....7..260F,*1969JMP....10..806L}, or to link meson and baryon masses in the quark model \cite{Ader:1981db,Nussinov:1999sx}. Before the applications to tetraquarks, we present  a brief review illustrated in the 3-body case, that follows the notation of \cite{Basdevant:1989pt,*Basdevant:1989pv,*Basdevant:1992cm,*1998FBS....24...39B,*2009FBS....46..199B}.

The \emph{naive} bound is deduced from the identity
\begin{equation}\label{eq:HP1}
\frac{\vec p_1^2+\vec p_2^2+\vec p_3^2}{2\,m}+\sum_{i<j} V_{ij}=\left[\frac{\vec p_1^2+\vec p_2^2}{4\,m}+V_{12}\right]+\cdots~,
\end{equation}
whose expectation value within the ground-state of the l.h.s.\ leads to the inequality 
\begin{equation}\label{eq:HP2}
 E_3(m,V)\ge 3\, E_2(2\,m,V)=\frac32\,E_2(m,2\,V)~,
\end{equation}
among the ground-state energies. For instance, in a simple additive quark model with a factor 1/2, i.e., $V=\sum_{i<j} v(r_{ij})/2$, with $v$ being the quarkonium potential, one gets 
$E_3(qqq)\ge 3\,E_2(q\bar q)/2$. This implies that a baryon is heavier per quark than a meson, as seen, e.g., by comparing $\Omega^-(1672)$ and $\phi(1020)$, of quark content $sss$ and $s\bar s$, respectively.

The inequality \eqref{eq:HP2} never becomes an equality as it contains unbalanced center-of-mass kinetic energy. If one starts instead from the intrinsic Hamiltonians, one gets saturation in the case of harmonic confinement. Namely
\begin{multline}\label{eq:HP3}
\frac{\vec p_1^2+\vec p_2^2+\vec p_3^2}{2\,m}-\frac{(\vec p_1+\vec p_2+\vec p_3)^2}{6\,m}+\sum_{i<j} V_{ij}\\
{}=\left[\frac{2}{3\,m}\,\genfrac{(}{)}{}{}{\vec p_2-\vec p_1}{2}^2+V_{12}\right]+\cdots~,
\end{multline}
leads to the \emph{improved} bound 
\begin{equation}\label{eq:HP4}
 E_3(m,V)\ge 3\, E_2(3\,m/2,V)~,
\end{equation}
which is better, as the energy $E_2$ is a decreasing function of the mass, for given $V$.

For unequal masses, this ''improved'' bound is straightforwardly generalized as (the potential terms are omitted)
\begin{eqnarray}\label{eq:HP5}
\sum_i \frac{\vec p_i^2}{2\,m_i}-\frac{(\sum_i \vec p_i)^2}{2\sum_i m_i}
{}&=&\left[\frac{1}{\mu_{12}}\,\genfrac{(}{)}{}{}{m_1\,\vec p_2-m_2\,\vec p_1}{m_1+m_2}^2
\right]+\cdots~,\nonumber\\
 E_3(m_1,m_2,m_3)&\ge& \sum_{i<j} E_2(\mu_{ij})~,\\
 \mu_{12}&=&2\,\frac{m_1\,m_2\sum_i m_i}{(m_1+m_2)^2}~.\nonumber
\end{eqnarray}
However, this inequality is not saturated for the harmonic oscillator. It can be improved by introducing a slightly  more general decomposition of the kinetic energy and optimizing some parameters. More precisely, this decomposition involves the parameters $b_i$, $y_i$ and $x_{ij}$ in the identity
\begin{multline}\label{eq:HP6}
\sum_i \frac{\vec p_i^2}{2\,m_i}=\left(\sum_i \vec p_i\right).\left(\sum_i b_i\,\vec p_i\right)\\
{}+\left[x_{12}^{-1}\,\genfrac{(}{)}{}{}{\vec p_2-y_3\,\vec p_1}{1+y_3}^2
\right]+\cdots
\end{multline}
For any given set $\{y_i\}$, one can determine the parameters $b_i$ and the  masses $x_{ij}$. If one takes the expectation value within the 3-body wave function, the first term of the r.h.s.\ disappears, and one reaches the so-called \emph{optimized} lower bound
\begin{equation}\label{eq:HP7}
 E_3\ge \max_{y_1,y_2,y_3}\sum_{i<j}E_2[x_{ij}(y_1,y_2,y_3)]~,
\end{equation}
where it can be shown that the maximization automatically fulfills $y_1\,y_2\,y_3=1$. 
\subsection{Application to tetraquarks}
Consider first the toy Hamiltonian \eqref{eq:H4}, slightly generalized as $r_{ij}\to v_{ij}=v(r_{ij})$ for all pairs. In the case of equal masses, which can be set to $m=M$  the simple identity
\begin{eqnarray}\label{eq:HP8}
 \sum_i \frac{\vec p_i^2}{2\,m}&+&\frac{v_{12}+v_{34}}{2}+\sum{\strut}'{\,}\frac{v_{ij}}{4}\nonumber\\
 {}&=&\frac{h_{12}(m)+h_{34}(m)}{2}+\sum{\strut}'{\,}\frac{h_{ij}(m)}{4}~,\\
h_{ij}(m)&=&\frac{\vec p_i^2+\vec p_j^2}{2\,m}+v_{ij}~,\nonumber
\end{eqnarray}
demonstrates that for the ground-states energies
\begin{equation}\label{eq:HP9}
 E_4(m)\ge 2\,E_2(m)~,
\end{equation}
i.e., the tetraquark with pure chromoelectric interaction and a frozen $\bar 33$ color wavefunction, is above twice the minimum of each $h_{ij}$, which is the threshold energy. This is the analog of the above ''naive'' lower bound. 

If one removes the center of mass, and starts from the decomposition
\begin{eqnarray}\label{eq:HP10}
\sum_i \frac{\vec p_i^2}{2\,m}&-&\frac{(\sum_i \vec p_i)^2}{8\,m}+\frac{v_{12}+v_{34}}{2}+\sum{\mathstrut}'{\,}\frac{v_{ij}}{4}\nonumber\\
{}&=&\frac{\tilde h_{12}(m)+\tilde h_{34}(m)}{2}+\sum{\strut}'{\,}\frac{\tilde h_{ij}(m/2)}{4}~,\\
 \tilde h_{ij}(m)&=&\frac1m\,\genfrac{(}{)}{}{}{\vec p_j-\vec p_i}{2}^2+v_{ij}~,\nonumber
 \end{eqnarray}
one gets the ''improved'' bound
\begin{equation}\label{eq:HP11}
 E_4(m)\ge E_2(m)+ E_2(m/2)~,
\end{equation}
that is better, as $E_2(m/2)>E_2(m)$. 
For unequal masses, the decomposition  reads
\begin{eqnarray}\label{eq:HP12}
\frac{\vec p_1^2+\vec p_2^2}{2\,M}&+&\frac{\vec p_3^2+\vec p_4^2}{2\,m}+\frac{v_{12}+v_{34}}{2}+\sum{\mathstrut}'{\,}\frac{v_{ij}}{4}\nonumber\\
{}&&\hspace{ -.5cm}=\left(\sum\vec p_i\right).(A(\vec p_1+\vec p_2)+B(\vec p_3+\vec p_4))\\
{}&+&\frac{\tilde h_{12}(x_{12})+\tilde h_{34}(x_{34})}{2}+\sum{\strut}'{\,}\frac{\tilde{\tilde h}_{ij}(x,a,b)}{4}~,\nonumber\\
\tilde{\tilde h}_{13}(x,a,b)&=&\frac1x\,\genfrac{(}{)}{}{}{\vec p_1-\vec p_3+ a\,\vec p_2+b\,\vec p_4}{2}^2+v_{ij}~,\nonumber
 \end{eqnarray}
where the masses $x_{12}$, $x_{34}$ and  $x$ are readily calculated from the parameters $A$, $B$ and $a$, and $b$.
This results into 
\begin{equation}\label{eq:HP13}
 E_4(M,m)\ge \max_{A,B,a,b}\left[E_2(x_{12})+ E_2(x_{34})+E_2(x)\right]~.
\end{equation}
Hence a rigorous lower bound is obtained from simple algebraic manipulations and the knowledge of the 2-body energy as a function of the reduced mass. For a linear interaction, \eqref{eq:HP13} further simplifies into
\begin{equation}\label{eq:HP14}
 E_4(M,m)\ge E_2(1)\,\max_{A,B,a,b}\left[x_{12}^{-1/3}+ x_{34}^{-1/3}+x^{-1/3}\right]~.
\end{equation}
where $E_2(1)=2.33811\ldots$ is the opposite of the first root of the Airy function. For $r^{0.1}$, the exponent $-1/3$ is replaced by $-0.1/2.1$ and $E_2(1)$ is computed numerically. The results for $E_4/E_2(1)$ as a function of $M/m$ are shown in Figs.~\ref{fig:HP1} and \ref{fig:HP01}. The sum $1/M+1/m$ is kept equal to 2 to fix the threshold energy at $2\,E_2(1)$. 
\section{Color mixing}\label{se:col-mix}
The $\ll{i}{j}$ model of Eq.~\eqref{eq:H4}, with a pairwise potential due to color-octet exchange, induces mixing between $\bar33$ and $6\bar6$ states in the $QQ-\bar q\bar q$ basis. Perhaps the true dynamics inhibits the call for higher color representations such as sextet, octet, etc., for the subsystems of a multiquarks, but for the time being, let us adopt the color-additive model. If one starts from a $\bar33$ state with $QQ$ in a spin triplet, and, for instance $\bar q\bar q=\bar u\bar d$ with spin and isospin $S=I=0$, then its orbital wave function is mainly made of an $s$-wave in all coordinates. It can mix with a color $6\bar 6$ with orbital excitations in the $\vec x$ and $\vec y$ linking $QQ$ and $\bar q\bar q$, respectively. A minimal wave function in this sector can be chosen as
\begin{eqnarray}
 \label{eq:psi6}
 \Psi_6 &&\propto \,\vec x.\vec y\,\exp(-a\,\vec x^2-b\,\vec y^2)~,\nonumber\\
 &&\qquad \text{or}\\
\Psi_6 &&\propto\, 
 \exp\bigl[-a_{12}\,\vec x^2-a_{34}\,\vec y^2 -\alpha(\vec r_{13}^2+\vec r_{24}^2)\nonumber\\
 &&\qquad\qquad{} -\beta(\vec r_{14}^2+\vec r_{23}^2)\bigr]-\{\alpha\leftrightarrow\beta\}~.\nonumber
\end{eqnarray}
The effect of color mixing for a spin-independent interaction was shown
 Fig.~\ref{fig:BilanL} in the case of a linear potential, and in Fig.~\ref{fig:BilanLC} for a Coulomb-plus-linear potential $V(r)=-a/r+b\,r$ with $a=0.4$, $b=0.2\,$GeV$^{2}$, and $m=0.3\,$GeV, as  function of $M/m$. The gain is less pronounced for very large $M/m$, but for the mass ratios of interest, color mixing is crucial to achieve binding. 

\begin{figure}[ht]
 \centering
\includegraphics[width=.35\textwidth]{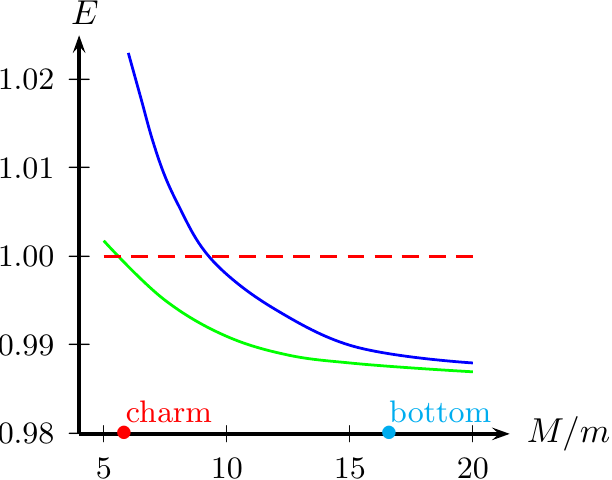}
 \caption{Effect of color-mixing on  the binding of $QQ\bar u\bar d$, within the AL1 model. The tetraquark energy is calculated with only the color $\bar33$ configurations (blue curve) and with the $6\bar6$ components (green curve). }
 \label{fig:AL1-col-mix}
\end{figure}

We now illustrate the role of color-mixing for the AL1 potential (to be introduced in Sec.~\ref{se:spin}).  The energy estimated as a function of $M/m$ without and with color-mixing is shown in Fig.~\ref{fig:AL1-col-mix}. The ground state of the $QQ\bar u\bar d$ that is candidate for stability, with $J^P=1^+$, has its main component with color $\bar33$, and spin $\{1,0\}$ in the $QQ-\bar u\bar d$ basis. The main admixture consists of $6\bar 6$ with spin $\{1,0\}$ and an antisymmetric orbital wavefunction of which \eqref{eq:psi6} is a prototype, and of $6\bar 6$ with spin $\{0,1\}$ with a symmetric orbital wavefunction.

The relevance of color-mixing has also been illustrated with realistic models in Table \rom{2} of Ref.~\cite{Vijande:2009zs} and has been stressed by several authors cited in \cite{Ballot:1983iv}, in particular Brink and Stancu.
\section{Spin-dependent corrections}\label{se:spin}
In the most advanced calculations of Ref.~\cite{Ballot:1983iv}, it was acknowledged that a pure additive interaction  such a \eqref{eq:H4} will not bind $cc\bar q\bar q$, on the sole basis that this tetraquark configuration benefits from the strong $cc$ chromoelectric attraction that is absent in the $Q\bar q+Q\bar q$ threshold. In the case where $qq=ud$, however, there is in addition a favorable chromomagnetic interaction in the tetraquark, while the threshold experiences only heavy-light spin-spin interaction, whose strength is suppressed by a factor $m/M$. 

\begin{figure}[ht]
 \centering
\includegraphics[width=.35\textwidth]{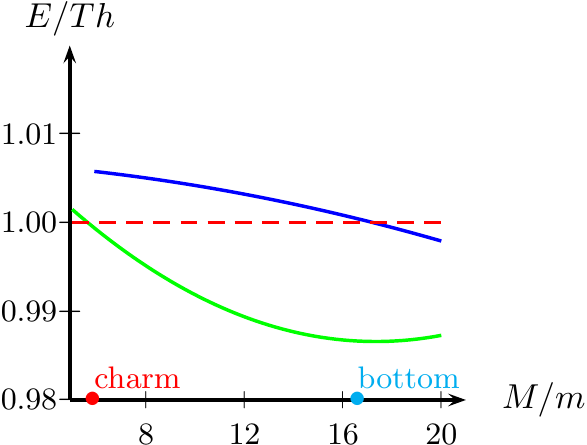}
 \caption{Effect of the spin-spin interaction of the binding of $QQ\bar u\bar d$, within the AL1 model. The tetraquark energy is calculated with (green line) and without (blue line) the chromomagnetic term. }
 \label{fig:AL1-SS}
\end{figure}

For illustration, we use the potential AL1 by Semay and Silvestre-Brac~\cite{Semay:1994ht}. Its central part is similar to the Coulomb-plus-linear adopted in Fig.~\ref{fig:BilanLC}. Its spin-spin part is a regularized Breit-Fermi interaction, with a smearing parameter that depends on the reduced mass. More precisely, 
\begin{gather}
\label{eq:V}
V_{ij}(r)= -\frac{\kappa}{r} +\lambda\,r-\Lambda
{}+ \frac{2\,\pi\,\alpha}{3\,m_i\,m_j}
\frac{\exp(-r^2/r_0^2)}{\pi^{3/2}\,r^3_0}\,\vec\sigma_i.
\vec\sigma_j~,\nonumber\\
r_0(m_i,m_j)=A\,\genfrac{(}{)}{}{0}{2\,m_i\,m_j}{m_i+m_j}^{-B}~,\\
m_q=0.315~,\quad m_c=1.836~,\quad m_b=5.227~,\ \nonumber\\
\Lambda=0.8321~,\quad B=0.2204~,\quad A=1.6553~,\nonumber \\
\kappa=0.5069~\quad \alpha=1.8609~,\quad \lambda=0.1653~,\nonumber
\end{gather}
where the units are appropriate powers of GeV. The results are shown in Fig.~\ref{fig:AL1-SS} for $QQ\bar u\bar d$, as a function of the mass ratio $M/m$.

The system $bb\bar u\bar d$ is barely bound without the spin-spin term, though the mass ratio $m_b/m_q$ is very large. Its acquires its binding energy of the order of 150\,MeV when the spin-spin is restored. 

The system $cc\bar u\bar d$ is clearly unbound when the spin-spin interaction is switched off. This is shown here for the AL1 model, but this is true for any realistic interaction, including an early model by Bhaduri et al.~\cite{Bhaduri:1981pn}. The case of $cc\bar u\bar d$ is actually remarkable. Here the binding requires both the color mixing of $\bar33$ with $6\bar 6$, and the spin-spin interaction. Moreover, the  binding is so tiny that it cannot be obtained with a simple variational method. One needs either a fully converged expansion on a basis of correlated Gaussians, or a hypersherical expansion up to a grand orbital momentum $K_\text{max}$ of the order of 12. 
Semay and Silvestre-Brac, who used their AL1 potential,  missed the binding, but their method of systematic expansion on the eigenstates of an harmonic oscillator is not very efficient to account for the short-range correlations, and is abandoned in the latest quark-model calculations. Janc and Rosina were the first to obtain binding with such potentials, and their calculation was checked by Barnea et al. (see \cite{Ballot:1983iv} for refs.).

\section{Conclusions}\label{se:concl}
Let us summarize. The four-body problem of tetraquarks is rather delicate, especially for systems at the edge of stability.
The analogy with atomic physics is a good guidance to indicate the most favorable configurations in the limit of dominant chromoelectric interaction. However, unlike the positronium molecule, the all-heavy configuration $QQ\bar Q\bar Q$ is not stable if one adopts a standard quark model and solves the four-body problem correctly. 

The method of Gaussian expansion works rather well. With most current models, the matrix elements can be estimated analytically and one can study the convergence as a function of the number of terms, and the role of each spin-color configuration entering a given tetraquark state.
This is also the case for the hyperspherical expansion.

The mixing of the $\bar 3 3$ and $6\bar 6$ color configurations is important, especially for states very near the threshold.  This mixing occurs by  both  the spin-independent and the spin-dependent parts of the potential. 

Approximations are welcome, especially if they shed some light on the four-body dynamics. The diquark-antidiquark approximation is not supported by a rigorous solution of the 4-body problem, but benefits of a stroke of luck, as the erroneous extra attraction introduced in the color $\bar 33$ channel is somewhat compensated by the neglect of the coupling to the color $6\bar 6$ channel. The equality relating $QQ\bar q\bar q$, $QQq$, $Qqq$ and $Q\bar q$ works surprisingly well as long as one is restricted to color $\bar 33$, but does not account for the attraction provided by color mixing. 

On the other hand, for asymmetric configurations $(QQ\bar q\bar q)$, the Born-Oppenheimer method provides a very good approximation, and an interesting insight into the dynamics. It has been probed here for a toy model with frozen color, and its extension as to include the coupling of color configurations would deserve some study. 

In short, $cc\bar u\bar d$ with $J^P=1^+$ is at the edge of binding within current quark models. For this state, all contributions to the binding should be added, in particular the mixing of states with different internal spin and color structure, and in addition, the four-body problem should be solved with extreme accuracy, for instance by pushing the hyperspherical expansion up to a grand angular momentum $K_\text{max}\ge 12$. 

In comparison, achieving the binding of $bb\bar u\bar d$ looks easier. Still, with a typical quark model, the stability of the ground state below the threshold cannot be reached if spin-effects and color mixing are  both neglected. The crucial role of spin effects explains why one does not expect too many states besides $1^+$~\cite{Vijande:2009kj}.

Needless to say that any improvement of the dynamics would be welcome. In \cite{Vijande:2009kj}, for example, this is done by including  some pion-exchange in the light quark sector. A better binding is obtained for $cc\bar u\bar d$. The presence of multi-body components in the interquark potential has been discussed, in particular a disconnected or connected string network linking the quarks and antiquarks. This string model provides an attraction that is larger than the pairwise linear interaction $\propto\,\sum \ll{i}{j}\,r_{ij}$, provided there is no constraint from the Pauli principle, i.e., that the color wave function can readjust itself freely when the quarks move. This is not the case for $cc\bar u\bar d$. A good test of that model would be the stability of flavor-asymmetric configurations such as $bc\bar u\bar s$. 

\vskip .1cm

\noindent\textsl{Note added:} the excess of attraction due to the point-like approximation for diquarks was also pointed out in \cite{Kiselev:2017eic} in the case of doubly-heavy baryons.
\acknowledgments
This work has been partially funded by Ministerio de Econom\'\i a, Industria y Competitividad
and EU FEDER under Contracts No.\ No. FPA2013- 47443, FPA2016-77177 and FPA2015-69714-REDT,
by Junta de Castilla y Le\'on under Contract No. SA041U16,
by Generalitat Valenciana PrometeoII/2014/066.
\clearpage

\begin{thebibliography}{54}%
\makeatletter
\providecommand \@ifxundefined [1]{%
 \@ifx{#1\undefined}
}%
\providecommand \@ifnum [1]{%
 \ifnum #1\expandafter \@firstoftwo
 \else \expandafter \@secondoftwo
 \fi
}%
\providecommand \@ifx [1]{%
 \ifx #1\expandafter \@firstoftwo
 \else \expandafter \@secondoftwo
 \fi
}%
\providecommand \natexlab [1]{#1}%
\providecommand \enquote  [1]{``#1''}%
\providecommand \bibnamefont  [1]{#1}%
\providecommand \bibfnamefont [1]{#1}%
\providecommand \citenamefont [1]{#1}%
\providecommand \href@noop [0]{\@secondoftwo}%
\providecommand \href [0]{\begingroup \@sanitize@url \@href}%
\providecommand \@href[1]{\@@startlink{#1}\@@href}%
\providecommand \@@href[1]{\endgroup#1\@@endlink}%
\providecommand \@sanitize@url [0]{\catcode `\\12\catcode `\$12\catcode
  `\&12\catcode `\#12\catcode `\^12\catcode `\_12\catcode `\%12\relax}%
\providecommand \@@startlink[1]{}%
\providecommand \@@endlink[0]{}%
\providecommand \url  [0]{\begingroup\@sanitize@url \@url }%
\providecommand \@url [1]{\endgroup\@href {#1}{\urlprefix }}%
\providecommand \urlprefix  [0]{URL }%
\providecommand \Eprint [0]{\href }%
\providecommand \doibase [0]{http://dx.doi.org/}%
\providecommand \selectlanguage [0]{\@gobble}%
\providecommand \bibinfo  [0]{\@secondoftwo}%
\providecommand \bibfield  [0]{\@secondoftwo}%
\providecommand \translation [1]{[#1]}%
\providecommand \BibitemOpen [0]{}%
\providecommand \bibitemStop [0]{}%
\providecommand \bibitemNoStop [0]{.\EOS\space}%
\providecommand \EOS [0]{\spacefactor3000\relax}%
\providecommand \BibitemShut  [1]{\csname bibitem#1\endcsname}%
\let\auto@bib@innerbib\@empty
\bibitem [{\citenamefont {Briceno}\ \emph {et~al.}(2016)\citenamefont {Briceno}
  \emph {et~al.}}]{Briceno:2015rlt}%
  \BibitemOpen
  \bibfield  {author} {\bibinfo {author} {\bibfnamefont {R.~A.}\ \bibnamefont
  {Briceno}} \emph {et~al.},\ }\bibfield  {title} {\enquote {\bibinfo {title}
  {{Issues and Opportunities in Exotic Hadrons}},}\ }\href {\doibase
  10.1088/1674-1137/40/4/042001} {\bibfield  {journal} {\bibinfo  {journal}
  {Chin. Phys.}\ }\textbf {\bibinfo {volume} {C40}},\ \bibinfo {pages} {042001}
  (\bibinfo {year} {2016})},\ \Eprint {http://arxiv.org/abs/1511.06779}
  {arXiv:1511.06779 [hep-ph]} \BibitemShut {NoStop}%
\bibitem [{\citenamefont {Lebed}\ \emph {et~al.}(2017)\citenamefont {Lebed},
  \citenamefont {Mitchell},\ and\ \citenamefont {Swanson}}]{Lebed:2016hpi}%
  \BibitemOpen
  \bibfield  {author} {\bibinfo {author} {\bibfnamefont {Richard~F.}\
  \bibnamefont {Lebed}}, \bibinfo {author} {\bibfnamefont {Ryan~E.}\
  \bibnamefont {Mitchell}}, \ and\ \bibinfo {author} {\bibfnamefont {Eric~S.}\
  \bibnamefont {Swanson}},\ }\bibfield  {title} {\enquote {\bibinfo {title}
  {{Heavy-Quark QCD Exotica}},}\ }\href {\doibase 10.1016/j.ppnp.2016.11.003}
  {\bibfield  {journal} {\bibinfo  {journal} {Prog. Part. Nucl. Phys.}\
  }\textbf {\bibinfo {volume} {93}},\ \bibinfo {pages} {143--194} (\bibinfo
  {year} {2017})},\ \Eprint {http://arxiv.org/abs/1610.04528} {arXiv:1610.04528
  [hep-ph]} \BibitemShut {NoStop}%
\bibitem [{\citenamefont {Chen}\ \emph {et~al.}(2016)\citenamefont {Chen},
  \citenamefont {Chen}, \citenamefont {Liu},\ and\ \citenamefont
  {Zhu}}]{Chen:2016qju}%
  \BibitemOpen
  \bibfield  {author} {\bibinfo {author} {\bibfnamefont {Hua-Xing}\
  \bibnamefont {Chen}}, \bibinfo {author} {\bibfnamefont {Wei}\ \bibnamefont
  {Chen}}, \bibinfo {author} {\bibfnamefont {Xiang}\ \bibnamefont {Liu}}, \
  and\ \bibinfo {author} {\bibfnamefont {Shi-Lin}\ \bibnamefont {Zhu}},\
  }\bibfield  {title} {\enquote {\bibinfo {title} {{The hidden-charm pentaquark
  and tetraquark states}},}\ }\href {\doibase 10.1016/j.physrep.2016.05.004} {\
   (\bibinfo {year} {2016}),\ 10.1016/j.physrep.2016.05.004},\ \Eprint
  {http://arxiv.org/abs/1601.02092} {arXiv:1601.02092 [hep-ph]} \BibitemShut
  {NoStop}%
\bibitem [{\citenamefont {Ali}\ \emph {et~al.}(2017)\citenamefont {Ali},
  \citenamefont {Lange},\ and\ \citenamefont {Stone}}]{Ali:2017jda}%
  \BibitemOpen
  \bibfield  {author} {\bibinfo {author} {\bibfnamefont {Ahmed}\ \bibnamefont
  {Ali}}, \bibinfo {author} {\bibfnamefont {Jens~Sören}\ \bibnamefont
  {Lange}}, \ and\ \bibinfo {author} {\bibfnamefont {Sheldon}\ \bibnamefont
  {Stone}},\ }\bibfield  {title} {\enquote {\bibinfo {title} {{Exotics: Heavy
  Pentaquarks and Tetraquarks}},}\ }\href@noop {} {\  (\bibinfo {year}
  {2017})},\ \Eprint {http://arxiv.org/abs/1706.00610} {arXiv:1706.00610
  [hep-ph]} \BibitemShut {NoStop}%
\bibitem [{\citenamefont {Esposito}\ \emph {et~al.}(2016)\citenamefont
  {Esposito}, \citenamefont {Pilloni},\ and\ \citenamefont
  {Polosa}}]{Esposito:2016noz}%
  \BibitemOpen
  \bibfield  {author} {\bibinfo {author} {\bibfnamefont {A.}~\bibnamefont
  {Esposito}}, \bibinfo {author} {\bibfnamefont {A.}~\bibnamefont {Pilloni}}, \
  and\ \bibinfo {author} {\bibfnamefont {A.~D.}\ \bibnamefont {Polosa}},\
  }\bibfield  {title} {\enquote {\bibinfo {title} {{Multiquark Resonances}},}\
  }\href {\doibase 10.1016/j.physrep.2016.11.002} {\bibfield  {journal}
  {\bibinfo  {journal} {Phys. Rept.}\ }\textbf {\bibinfo {volume} {668}},\
  \bibinfo {pages} {1--97} (\bibinfo {year} {2016})},\ \Eprint
  {http://arxiv.org/abs/1611.07920} {arXiv:1611.07920 [hep-ph]} \BibitemShut
  {NoStop}%
\bibitem [{\citenamefont {Richard}(2016)}]{Richard:2016eis}%
  \BibitemOpen
  \bibfield  {author} {\bibinfo {author} {\bibfnamefont {Jean-Marc}\
  \bibnamefont {Richard}},\ }\bibfield  {title} {\enquote {\bibinfo {title}
  {{Exotic hadrons: review and perspectives}},}\ }\href {\doibase
  10.1007/s00601-016-1159-0} {\bibfield  {journal} {\bibinfo  {journal} {Few
  Body Syst.}\ }\textbf {\bibinfo {volume} {57}},\ \bibinfo {pages}
  {1185--1212} (\bibinfo {year} {2016})},\ \bibinfo {note} {{Special issue for
  the 30th anniversary of Few-Body Systems}},\ \Eprint
  {http://arxiv.org/abs/1606.08593} {arXiv:1606.08593 [hep-ph]} \BibitemShut
  {NoStop}%
\bibitem [{\citenamefont {Aaij}\ \emph {et~al.}(2017)\citenamefont {Aaij} \emph
  {et~al.}}]{Aaij:2017ueg}%
  \BibitemOpen
  \bibfield  {author} {\bibinfo {author} {\bibfnamefont {Roel}\ \bibnamefont
  {Aaij}} \emph {et~al.} (\bibinfo {collaboration} {LHCb}),\ }\bibfield
  {title} {\enquote {\bibinfo {title} {{Observation of the doubly charmed
  baryon $\Xi_{cc}^{++}$}},}\ }\href {\doibase 10.1103/PhysRevLett.119.112001}
  {\bibfield  {journal} {\bibinfo  {journal} {Phys. Rev. Lett.}\ }\textbf
  {\bibinfo {volume} {119}},\ \bibinfo {pages} {112001} (\bibinfo {year}
  {2017})},\ \Eprint {http://arxiv.org/abs/1707.01621} {arXiv:1707.01621
  [hep-ex]} \BibitemShut {NoStop}%
\bibitem [{\citenamefont {Ader}\ \emph {et~al.}(1982)\citenamefont {Ader},
  \citenamefont {Richard},\ and\ \citenamefont {Taxil}}]{Ader:1981db}%
  \BibitemOpen
  \bibfield  {author} {\bibinfo {author} {\bibfnamefont {J.~P.}\ \bibnamefont
  {Ader}}, \bibinfo {author} {\bibfnamefont {J.~M.}\ \bibnamefont {Richard}}, \
  and\ \bibinfo {author} {\bibfnamefont {P.}~\bibnamefont {Taxil}},\ }\bibfield
   {title} {\enquote {\bibinfo {title} {Do narrow heavy multiquark states
  exist?}}\ }\href@noop {} {\bibfield  {journal} {\bibinfo  {journal} {Phys.
  Rev.}\ }\textbf {\bibinfo {volume} {D25}},\ \bibinfo {pages} {2370} (\bibinfo
  {year} {1982})}\BibitemShut {NoStop}%
\bibitem [{\citenamefont {Ballot}\ and\ \citenamefont
  {Richard}(1983)}]{Ballot:1983iv}%
  \BibitemOpen
  \bibfield  {author} {\bibinfo {author} {\bibfnamefont {J.~l.}\ \bibnamefont
  {Ballot}}\ and\ \bibinfo {author} {\bibfnamefont {J.~M.}\ \bibnamefont
  {Richard}},\ }\bibfield  {title} {\enquote {\bibinfo {title}
  {{F\lowercase{OUR QUARK STATES IN ADDITIVE POTENTIALS}}},}\ }\href {\doibase
  10.1016/0370-2693(83)90991-7} {\bibfield  {journal} {\bibinfo  {journal}
  {Phys. Lett.}\ }\textbf {\bibinfo {volume} {123B}},\ \bibinfo {pages}
  {449--451} (\bibinfo {year} {1983})}\BibitemShut {NoStop}%
\bibitem [{\citenamefont {Zouzou}\ \emph {et~al.}(1986)\citenamefont {Zouzou},
  \citenamefont {Silvestre-Brac}, \citenamefont {Gignoux},\ and\ \citenamefont
  {Richard}}]{Zouzou:1986qh}%
  \BibitemOpen
  \bibfield  {author} {\bibinfo {author} {\bibfnamefont {S.}~\bibnamefont
  {Zouzou}}, \bibinfo {author} {\bibfnamefont {B.}~\bibnamefont
  {Silvestre-Brac}}, \bibinfo {author} {\bibfnamefont {C.}~\bibnamefont
  {Gignoux}}, \ and\ \bibinfo {author} {\bibfnamefont {J.~M.}\ \bibnamefont
  {Richard}},\ }\bibfield  {title} {\enquote {\bibinfo {title} {{Four quark
  bound states}},}\ }\href {\doibase 10.1007/BF01557611} {\bibfield  {journal}
  {\bibinfo  {journal} {Z. Phys.}\ }\textbf {\bibinfo {volume} {C30}},\
  \bibinfo {pages} {457} (\bibinfo {year} {1986})}\BibitemShut {NoStop}%
\bibitem [{\citenamefont {Heller}\ and\ \citenamefont
  {Tjon}(1985)}]{Heller:1985cb}%
  \BibitemOpen
  \bibfield  {author} {\bibinfo {author} {\bibfnamefont {L.}~\bibnamefont
  {Heller}}\ and\ \bibinfo {author} {\bibfnamefont {J.~A.}\ \bibnamefont
  {Tjon}},\ }\bibfield  {title} {\enquote {\bibinfo {title} {{On Bound States
  of Heavy $Q^2 \bar{Q}^2$ Systems}},}\ }\href {\doibase
  10.1103/PhysRevD.32.755} {\bibfield  {journal} {\bibinfo  {journal} {Phys.
  Rev.}\ }\textbf {\bibinfo {volume} {D32}},\ \bibinfo {pages} {755} (\bibinfo
  {year} {1985})}\BibitemShut {NoStop}%
\bibitem [{\citenamefont {Carlson}\ \emph {et~al.}(1988)\citenamefont
  {Carlson}, \citenamefont {Heller},\ and\ \citenamefont
  {Tjon}}]{Carlson:1988hh}%
  \BibitemOpen
  \bibfield  {author} {\bibinfo {author} {\bibfnamefont {J.}~\bibnamefont
  {Carlson}}, \bibinfo {author} {\bibfnamefont {L.}~\bibnamefont {Heller}}, \
  and\ \bibinfo {author} {\bibfnamefont {J.~A.}\ \bibnamefont {Tjon}},\
  }\bibfield  {title} {\enquote {\bibinfo {title} {S\lowercase{TABILITY OF
  DIMESONS}},}\ }\href@noop {} {\bibfield  {journal} {\bibinfo  {journal}
  {Phys. Rev.}\ }\textbf {\bibinfo {volume} {D37}},\ \bibinfo {pages} {744}
  (\bibinfo {year} {1988})}\BibitemShut {NoStop}%
\bibitem [{\citenamefont {Heller}\ and\ \citenamefont
  {Tjon}(1987)}]{Heller:1986bt}%
  \BibitemOpen
  \bibfield  {author} {\bibinfo {author} {\bibfnamefont {L.}~\bibnamefont
  {Heller}}\ and\ \bibinfo {author} {\bibfnamefont {J.~A.}\ \bibnamefont
  {Tjon}},\ }\bibfield  {title} {\enquote {\bibinfo {title} {O\lowercase{N THE
  EXISTENCE OF STABLE DIMESONS}},}\ }\href@noop {} {\bibfield  {journal}
  {\bibinfo  {journal} {Phys. Rev.}\ }\textbf {\bibinfo {volume} {D35}},\
  \bibinfo {pages} {969} (\bibinfo {year} {1987})}\BibitemShut {NoStop}%
\bibitem [{\citenamefont {Brink}\ and\ \citenamefont
  {Stancu}(1994)}]{Brink:1994ic}%
  \BibitemOpen
  \bibfield  {author} {\bibinfo {author} {\bibfnamefont {D.~M.}\ \bibnamefont
  {Brink}}\ and\ \bibinfo {author} {\bibfnamefont {F.}~\bibnamefont {Stancu}},\
  }\bibfield  {title} {\enquote {\bibinfo {title} {{Role of hidden color states
  in $qq\bar q\bar q$ systems}},}\ }\href {\doibase 10.1103/PhysRevD.49.4665}
  {\bibfield  {journal} {\bibinfo  {journal} {Phys. Rev.}\ }\textbf {\bibinfo
  {volume} {D49}},\ \bibinfo {pages} {4665--4674} (\bibinfo {year}
  {1994})}\BibitemShut {NoStop}%
\bibitem [{\citenamefont {Brink}\ and\ \citenamefont
  {Stancu}(1998)}]{Brink:1998as}%
  \BibitemOpen
  \bibfield  {author} {\bibinfo {author} {\bibfnamefont {D.~M.}\ \bibnamefont
  {Brink}}\ and\ \bibinfo {author} {\bibfnamefont {Fl.}\ \bibnamefont
  {Stancu}},\ }\bibfield  {title} {\enquote {\bibinfo {title} {{Tetraquarks
  with heavy flavors}},}\ }\href {\doibase 10.1103/PhysRevD.57.6778} {\bibfield
   {journal} {\bibinfo  {journal} {Phys. Rev.}\ }\textbf {\bibinfo {volume}
  {D57}},\ \bibinfo {pages} {6778--6787} (\bibinfo {year} {1998})}\BibitemShut
  {NoStop}%
\bibitem [{\citenamefont {Vijande}\ \emph {et~al.}(2004)\citenamefont
  {Vijande}, \citenamefont {Fernandez}, \citenamefont {Valcarce},\ and\
  \citenamefont {Silvestre-Brac}}]{Vijande:2003ki}%
  \BibitemOpen
  \bibfield  {author} {\bibinfo {author} {\bibfnamefont {J.}~\bibnamefont
  {Vijande}}, \bibinfo {author} {\bibfnamefont {F.}~\bibnamefont {Fernandez}},
  \bibinfo {author} {\bibfnamefont {A.}~\bibnamefont {Valcarce}}, \ and\
  \bibinfo {author} {\bibfnamefont {B.}~\bibnamefont {Silvestre-Brac}},\
  }\bibfield  {title} {\enquote {\bibinfo {title} {{Tetraquarks in a chiral
  constituent quark model}},}\ }\href {\doibase 10.1140/epja/i2003-10128-9}
  {\bibfield  {journal} {\bibinfo  {journal} {Eur. Phys. J.}\ }\textbf
  {\bibinfo {volume} {A19}},\ \bibinfo {pages} {383} (\bibinfo {year}
  {2004})},\ \Eprint {http://arxiv.org/abs/hep-ph/0310007}
  {arXiv:hep-ph/0310007 [hep-ph]} \BibitemShut {NoStop}%
\bibitem [{\citenamefont {Janc}\ and\ \citenamefont
  {Rosina}(2004)}]{Janc:2004qn}%
  \BibitemOpen
  \bibfield  {author} {\bibinfo {author} {\bibfnamefont {D.}~\bibnamefont
  {Janc}}\ and\ \bibinfo {author} {\bibfnamefont {M.}~\bibnamefont {Rosina}},\
  }\bibfield  {title} {\enquote {\bibinfo {title} {{The $T_{cc}= D D^*$
  molecular state}},}\ }\href@noop {} {\bibfield  {journal} {\bibinfo
  {journal} {Few Body Syst.}\ }\textbf {\bibinfo {volume} {35}},\ \bibinfo
  {pages} {175--196} (\bibinfo {year} {2004})},\ \Eprint
  {http://arxiv.org/abs/hep-ph/0405208} {hep-ph/0405208} \BibitemShut {NoStop}%
\bibitem [{\citenamefont {Vijande}\ \emph
  {et~al.}(2007{\natexlab{a}})\citenamefont {Vijande}, \citenamefont
  {Weissman}, \citenamefont {Valcarce},\ and\ \citenamefont
  {Barnea}}]{Vijande:2007rf}%
  \BibitemOpen
  \bibfield  {author} {\bibinfo {author} {\bibfnamefont {J.}~\bibnamefont
  {Vijande}}, \bibinfo {author} {\bibfnamefont {E.}~\bibnamefont {Weissman}},
  \bibinfo {author} {\bibfnamefont {A.}~\bibnamefont {Valcarce}}, \ and\
  \bibinfo {author} {\bibfnamefont {N.}~\bibnamefont {Barnea}},\ }\bibfield
  {title} {\enquote {\bibinfo {title} {{Are there compact heavy four-quark
  bound states?}}}\ }\href {\doibase 10.1103/PhysRevD.76.094027} {\bibfield
  {journal} {\bibinfo  {journal} {Phys. Rev.}\ }\textbf {\bibinfo {volume}
  {D76}},\ \bibinfo {pages} {094027} (\bibinfo {year} {2007}{\natexlab{a}})},\
  \Eprint {http://arxiv.org/abs/0710.2516} {arXiv:0710.2516 [hep-ph]}
  \BibitemShut {NoStop}%
\bibitem [{\citenamefont {Vijande}\ \emph
  {et~al.}(2007{\natexlab{b}})\citenamefont {Vijande}, \citenamefont
  {Valcarce},\ and\ \citenamefont {Richard}}]{Vijande:2007ix}%
  \BibitemOpen
  \bibfield  {author} {\bibinfo {author} {\bibfnamefont {J.}~\bibnamefont
  {Vijande}}, \bibinfo {author} {\bibfnamefont {A.}~\bibnamefont {Valcarce}}, \
  and\ \bibinfo {author} {\bibfnamefont {J.~M.}\ \bibnamefont {Richard}},\
  }\bibfield  {title} {\enquote {\bibinfo {title} {{Stability of multiquarks in
  a simple string model}},}\ }\href {\doibase 10.1103/PhysRevD.76.114013}
  {\bibfield  {journal} {\bibinfo  {journal} {Phys. Rev.}\ }\textbf {\bibinfo
  {volume} {D76}},\ \bibinfo {pages} {114013} (\bibinfo {year}
  {2007}{\natexlab{b}})},\ \Eprint {http://arxiv.org/abs/0707.3996}
  {arXiv:0707.3996 [hep-ph]} \BibitemShut {NoStop}%
\bibitem [{\citenamefont {Carames}\ \emph {et~al.}(2011)\citenamefont
  {Carames}, \citenamefont {Valcarce},\ and\ \citenamefont
  {Vijande}}]{Carames:2011zz}%
  \BibitemOpen
  \bibfield  {author} {\bibinfo {author} {\bibfnamefont {T.~F.}\ \bibnamefont
  {Carames}}, \bibinfo {author} {\bibfnamefont {A.}~\bibnamefont {Valcarce}}, \
  and\ \bibinfo {author} {\bibfnamefont {J.}~\bibnamefont {Vijande}},\
  }\bibfield  {title} {\enquote {\bibinfo {title} {{Doubly charmed exotic
  mesons: A gift of nature?}}}\ }\href {\doibase
  10.1016/j.physletb.2011.04.023} {\bibfield  {journal} {\bibinfo  {journal}
  {Phys. Lett.}\ }\textbf {\bibinfo {volume} {B699}},\ \bibinfo {pages}
  {291--295} (\bibinfo {year} {2011})}\BibitemShut {NoStop}%
\bibitem [{\citenamefont {Hyodo}\ \emph {et~al.}(2013)\citenamefont {Hyodo},
  \citenamefont {Liu}, \citenamefont {Oka}, \citenamefont {Sudoh},\ and\
  \citenamefont {Yasui}}]{Hyodo:2012pm}%
  \BibitemOpen
  \bibfield  {author} {\bibinfo {author} {\bibfnamefont {Tetsuo}\ \bibnamefont
  {Hyodo}}, \bibinfo {author} {\bibfnamefont {Yan-Rui}\ \bibnamefont {Liu}},
  \bibinfo {author} {\bibfnamefont {Makoto}\ \bibnamefont {Oka}}, \bibinfo
  {author} {\bibfnamefont {Kazutaka}\ \bibnamefont {Sudoh}}, \ and\ \bibinfo
  {author} {\bibfnamefont {Shigehiro}\ \bibnamefont {Yasui}},\ }\bibfield
  {title} {\enquote {\bibinfo {title} {{Production of doubly charmed
  tetraquarks with exotic color configurations in electron-positron
  collisions}},}\ }\href {\doibase 10.1016/j.physletb.2013.02.045} {\bibfield
  {journal} {\bibinfo  {journal} {Phys. Lett.}\ }\textbf {\bibinfo {volume}
  {B721}},\ \bibinfo {pages} {56--60} (\bibinfo {year} {2013})},\ \Eprint
  {http://arxiv.org/abs/1209.6207} {arXiv:1209.6207 [hep-ph]} \BibitemShut
  {NoStop}%
\bibitem [{\citenamefont {Mehen}(2017)}]{Mehen:2017nrh}%
  \BibitemOpen
  \bibfield  {author} {\bibinfo {author} {\bibfnamefont {Thomas}\ \bibnamefont
  {Mehen}},\ }\bibfield  {title} {\enquote {\bibinfo {title} {{Implications of
  Heavy Quark-Diquark Symmetry for Excited Doubly Heavy Baryons and
  Tetraquarks}},}\ }\href {\doibase 10.1103/PhysRevD.96.094028} {\bibfield
  {journal} {\bibinfo  {journal} {Phys. Rev.}\ }\textbf {\bibinfo {volume}
  {D96}},\ \bibinfo {pages} {094028} (\bibinfo {year} {2017})},\ \Eprint
  {http://arxiv.org/abs/1708.05020} {arXiv:1708.05020 [hep-ph]} \BibitemShut
  {NoStop}%
\bibitem [{\citenamefont {Yasui}\ \emph {et~al.}(2013)\citenamefont {Yasui},
  \citenamefont {Ohkoda}, \citenamefont {Yamaguchi}, \citenamefont {Sudoh},\
  and\ \citenamefont {Hosaka}}]{Yasui:2013tsa}%
  \BibitemOpen
  \bibfield  {author} {\bibinfo {author} {\bibfnamefont {S.}~\bibnamefont
  {Yasui}}, \bibinfo {author} {\bibfnamefont {S.}~\bibnamefont {Ohkoda}},
  \bibinfo {author} {\bibfnamefont {Y.}~\bibnamefont {Yamaguchi}}, \bibinfo
  {author} {\bibfnamefont {K.}~\bibnamefont {Sudoh}}, \ and\ \bibinfo {author}
  {\bibfnamefont {A.}~\bibnamefont {Hosaka}},\ }\bibfield  {title} {\enquote
  {\bibinfo {title} {{Doubly Charmed Exotic Mesons}},}\ }\bibfield  {booktitle}
  {\emph {\bibinfo {booktitle} {{Proceedings, 20th International IUPAP
  Conference on Few-Body Problems in Physics (FB20): Fukuoka, Japan, August
  20-25, 2012}}},\ }\href {\doibase 10.1007/s00601-013-0629-x} {\bibfield
  {journal} {\bibinfo  {journal} {Few Body Syst.}\ }\textbf {\bibinfo {volume}
  {54}},\ \bibinfo {pages} {1023--1026} (\bibinfo {year} {2013})}\BibitemShut
  {NoStop}%
\bibitem [{\citenamefont {Czarnecki}\ \emph {et~al.}(2018)\citenamefont
  {Czarnecki}, \citenamefont {Leng},\ and\ \citenamefont
  {Voloshin}}]{Czarnecki:2017vco}%
  \BibitemOpen
  \bibfield  {author} {\bibinfo {author} {\bibfnamefont {Andrzej}\ \bibnamefont
  {Czarnecki}}, \bibinfo {author} {\bibfnamefont {Bo}~\bibnamefont {Leng}}, \
  and\ \bibinfo {author} {\bibfnamefont {M.~B.}\ \bibnamefont {Voloshin}},\
  }\bibfield  {title} {\enquote {\bibinfo {title} {{Stability of tetrons}},}\
  }\href {\doibase 10.1016/j.physletb.2018.01.034} {\bibfield  {journal}
  {\bibinfo  {journal} {Phys. Lett.}\ }\textbf {\bibinfo {volume} {B778}},\
  \bibinfo {pages} {233--238} (\bibinfo {year} {2018})},\ \Eprint
  {http://arxiv.org/abs/1708.04594} {arXiv:1708.04594 [hep-ph]} \BibitemShut
  {NoStop}%
\bibitem [{\citenamefont {Richard}\ \emph {et~al.}(2017)\citenamefont
  {Richard}, \citenamefont {Valcarce},\ and\ \citenamefont
  {Vijande}}]{Richard:2017vry}%
  \BibitemOpen
  \bibfield  {author} {\bibinfo {author} {\bibfnamefont {Jean-Marc}\
  \bibnamefont {Richard}}, \bibinfo {author} {\bibfnamefont {Alfredo}\
  \bibnamefont {Valcarce}}, \ and\ \bibinfo {author} {\bibfnamefont {Javier}\
  \bibnamefont {Vijande}},\ }\bibfield  {title} {\enquote {\bibinfo {title}
  {{String dynamics and metastability of all-heavy tetraquarks}},}\ }\href
  {\doibase 10.1103/PhysRevD.95.054019} {\bibfield  {journal} {\bibinfo
  {journal} {Phys. Rev.}\ }\textbf {\bibinfo {volume} {D95}},\ \bibinfo {pages}
  {054019} (\bibinfo {year} {2017})},\ \Eprint
  {http://arxiv.org/abs/1703.00783} {arXiv:1703.00783 [hep-ph]} \BibitemShut
  {NoStop}%
\bibitem [{\citenamefont {Vijande}\ and\ \citenamefont
  {Valcarce}(2009)}]{Vijande:2009zs}%
  \BibitemOpen
  \bibfield  {author} {\bibinfo {author} {\bibfnamefont {J.}~\bibnamefont
  {Vijande}}\ and\ \bibinfo {author} {\bibfnamefont {A.}~\bibnamefont
  {Valcarce}},\ }\bibfield  {title} {\enquote {\bibinfo {title} {{Probabilities
  in nonorthogonal basis: Four- quark systems}},}\ }\href {\doibase
  10.1103/PhysRevC.80.035204} {\bibfield  {journal} {\bibinfo  {journal} {Phys.
  Rev.}\ }\textbf {\bibinfo {volume} {C80}},\ \bibinfo {pages} {035204}
  (\bibinfo {year} {2009})},\ \Eprint {http://arxiv.org/abs/0908.3254}
  {arXiv:0908.3254 [hep-ph]} \BibitemShut {NoStop}%
\bibitem [{\citenamefont {{Bethe}}\ and\ \citenamefont
  {{Salpeter}}(1957)}]{1957qmot.book.....B}%
  \BibitemOpen
  \bibfield  {author} {\bibinfo {author} {\bibfnamefont {H.~A.}\ \bibnamefont
  {{Bethe}}}\ and\ \bibinfo {author} {\bibfnamefont {E.~E.}\ \bibnamefont
  {{Salpeter}}},\ }\href@noop {} {\emph {\bibinfo {title} {Quantum Mechanics of
  One- and Two-Electron Atoms, New York: Academic Press, 1957}}}\ (\bibinfo
  {year} {1957})\BibitemShut {NoStop}%
\bibitem [{\citenamefont {{H{\o}gaasen}}\ \emph {et~al.}(2010)\citenamefont
  {{H{\o}gaasen}}, \citenamefont {{Richard}},\ and\ \citenamefont
  {{Sorba}}}]{2010AmJPh..78...86H}%
  \BibitemOpen
  \bibfield  {author} {\bibinfo {author} {\bibfnamefont {H.}~\bibnamefont
  {{H{\o}gaasen}}}, \bibinfo {author} {\bibfnamefont {J.-M.}\ \bibnamefont
  {{Richard}}}, \ and\ \bibinfo {author} {\bibfnamefont {P.}~\bibnamefont
  {{Sorba}}},\ }\bibfield  {title} {\enquote {\bibinfo {title} {{Two-electron
  atoms, ions, and molecules}},}\ }\href {\doibase 10.1119/1.3236392}
  {\bibfield  {journal} {\bibinfo  {journal} {American Journal of Physics}\
  }\textbf {\bibinfo {volume} {78}},\ \bibinfo {pages} {86--93} (\bibinfo
  {year} {2010})},\ \Eprint {http://arxiv.org/abs/0907.2614} {arXiv:0907.2614
  [quant-ph]} \BibitemShut {NoStop}%
\bibitem [{\citenamefont {Hiyama}\ \emph {et~al.}(2003)\citenamefont {Hiyama},
  \citenamefont {Kino},\ and\ \citenamefont {Kamimura}}]{Hiyama:2003cu}%
  \BibitemOpen
  \bibfield  {author} {\bibinfo {author} {\bibfnamefont {E.}~\bibnamefont
  {Hiyama}}, \bibinfo {author} {\bibfnamefont {Y.}~\bibnamefont {Kino}}, \ and\
  \bibinfo {author} {\bibfnamefont {M.}~\bibnamefont {Kamimura}},\ }\bibfield
  {title} {\enquote {\bibinfo {title} {{Gaussian expansion method for few-body
  systems}},}\ }\href {\doibase 10.1016/S0146-6410(03)90015-9} {\bibfield
  {journal} {\bibinfo  {journal} {Prog. Part. Nucl. Phys.}\ }\textbf {\bibinfo
  {volume} {51}},\ \bibinfo {pages} {223--307} (\bibinfo {year}
  {2003})}\BibitemShut {NoStop}%
\bibitem [{\citenamefont {{Mitroy}}\ \emph {et~al.}(2013)\citenamefont
  {{Mitroy}}, \citenamefont {{Bubin}}, \citenamefont {{Horiuchi}},
  \citenamefont {{Suzuki}}, \citenamefont {{Adamowicz}}, \citenamefont
  {{Cencek}}, \citenamefont {{Szalewicz}}, \citenamefont {{Komasa}},
  \citenamefont {{Blume}},\ and\ \citenamefont
  {{Varga}}}]{2013RvMP...85..693M}%
  \BibitemOpen
  \bibfield  {author} {\bibinfo {author} {\bibfnamefont {J.}~\bibnamefont
  {{Mitroy}}}, \bibinfo {author} {\bibfnamefont {S.}~\bibnamefont {{Bubin}}},
  \bibinfo {author} {\bibfnamefont {W.}~\bibnamefont {{Horiuchi}}}, \bibinfo
  {author} {\bibfnamefont {Y.}~\bibnamefont {{Suzuki}}}, \bibinfo {author}
  {\bibfnamefont {L.}~\bibnamefont {{Adamowicz}}}, \bibinfo {author}
  {\bibfnamefont {W.}~\bibnamefont {{Cencek}}}, \bibinfo {author}
  {\bibfnamefont {K.}~\bibnamefont {{Szalewicz}}}, \bibinfo {author}
  {\bibfnamefont {J.}~\bibnamefont {{Komasa}}}, \bibinfo {author}
  {\bibfnamefont {D.}~\bibnamefont {{Blume}}}, \ and\ \bibinfo {author}
  {\bibfnamefont {K.}~\bibnamefont {{Varga}}},\ }\bibfield  {title} {\enquote
  {\bibinfo {title} {{Theory and application of explicitly correlated
  Gaussians}},}\ }\href {\doibase 10.1103/RevModPhys.85.693} {\bibfield
  {journal} {\bibinfo  {journal} {Reviews of Modern Physics}\ }\textbf
  {\bibinfo {volume} {85}},\ \bibinfo {pages} {693--749} (\bibinfo {year}
  {2013})}\BibitemShut {NoStop}%
\bibitem [{\citenamefont {Vijande}\ \emph {et~al.}(2009)\citenamefont
  {Vijande}, \citenamefont {Valcarce},\ and\ \citenamefont
  {Barnea}}]{Vijande:2009kj}%
  \BibitemOpen
  \bibfield  {author} {\bibinfo {author} {\bibfnamefont {J.}~\bibnamefont
  {Vijande}}, \bibinfo {author} {\bibfnamefont {A.}~\bibnamefont {Valcarce}}, \
  and\ \bibinfo {author} {\bibfnamefont {N.}~\bibnamefont {Barnea}},\
  }\bibfield  {title} {\enquote {\bibinfo {title} {{Exotic meson-meson
  molecules and compact four--quark states}},}\ }\href {\doibase
  10.1103/PhysRevD.79.074010} {\bibfield  {journal} {\bibinfo  {journal} {Phys.
  Rev.}\ }\textbf {\bibinfo {volume} {D79}},\ \bibinfo {pages} {074010}
  (\bibinfo {year} {2009})},\ \Eprint {http://arxiv.org/abs/0903.2949}
  {arXiv:0903.2949 [hep-ph]} \BibitemShut {NoStop}%
\bibitem [{\citenamefont {Anselmino}\ \emph {et~al.}(1993)\citenamefont
  {Anselmino}, \citenamefont {Predazzi}, \citenamefont {Ekelin}, \citenamefont
  {Fredriksson},\ and\ \citenamefont {Lichtenberg}}]{Anselmino:1992vg}%
  \BibitemOpen
  \bibfield  {author} {\bibinfo {author} {\bibfnamefont {Mauro}\ \bibnamefont
  {Anselmino}}, \bibinfo {author} {\bibfnamefont {Enrico}\ \bibnamefont
  {Predazzi}}, \bibinfo {author} {\bibfnamefont {Svante}\ \bibnamefont
  {Ekelin}}, \bibinfo {author} {\bibfnamefont {Sverker}\ \bibnamefont
  {Fredriksson}}, \ and\ \bibinfo {author} {\bibfnamefont {D.~B.}\ \bibnamefont
  {Lichtenberg}},\ }\bibfield  {title} {\enquote {\bibinfo {title}
  {{Diquarks}},}\ }\href {\doibase 10.1103/RevModPhys.65.1199} {\bibfield
  {journal} {\bibinfo  {journal} {Rev. Mod. Phys.}\ }\textbf {\bibinfo {volume}
  {65}},\ \bibinfo {pages} {1199--1234} (\bibinfo {year} {1993})}\BibitemShut
  {NoStop}%
\bibitem [{\citenamefont {Klempt}\ \emph {et~al.}(2017)\citenamefont {Klempt},
  \citenamefont {Sarantsev},\ and\ \citenamefont {Thoma}}]{Klempt:2017lwq}%
  \BibitemOpen
  \bibfield  {author} {\bibinfo {author} {\bibfnamefont {Eberhard}\
  \bibnamefont {Klempt}}, \bibinfo {author} {\bibfnamefont {Andrey~V.}\
  \bibnamefont {Sarantsev}}, \ and\ \bibinfo {author} {\bibfnamefont {Ulrike}\
  \bibnamefont {Thoma}},\ }\bibfield  {title} {\enquote {\bibinfo {title}
  {{Partial wave analysis}},}\ }\bibfield  {booktitle} {\emph {\bibinfo
  {booktitle} {{Proceedings, SFB/TRR16 Symposium: Subnuclear Structure of
  Matter: Achievements and Challenges: Bonn, Germany, June 6-9, 2016}}},\
  }\href {\doibase 10.1051/epjconf/201713402002} {\bibfield  {journal}
  {\bibinfo  {journal} {EPJ Web Conf.}\ }\textbf {\bibinfo {volume} {134}},\
  \bibinfo {pages} {02002} (\bibinfo {year} {2017})}\BibitemShut {NoStop}%
\bibitem [{\citenamefont {Fredriksson}\ and\ \citenamefont
  {Jandel}(1982)}]{Fredriksson:1981mh}%
  \BibitemOpen
  \bibfield  {author} {\bibinfo {author} {\bibfnamefont {Sverker}\ \bibnamefont
  {Fredriksson}}\ and\ \bibinfo {author} {\bibfnamefont {Magnus}\ \bibnamefont
  {Jandel}},\ }\bibfield  {title} {\enquote {\bibinfo {title} {{The Diquark
  Deuteron}},}\ }\href {\doibase 10.1103/PhysRevLett.48.14} {\bibfield
  {journal} {\bibinfo  {journal} {Phys. Rev. Lett.}\ }\textbf {\bibinfo
  {volume} {48}},\ \bibinfo {pages} {14} (\bibinfo {year} {1982})}\BibitemShut
  {NoStop}%
\bibitem [{\citenamefont {Jaffe}(1986)}]{Jaffee:1986zz}%
  \BibitemOpen
  \bibfield  {author} {\bibinfo {author} {\bibfnamefont {R.~L.}\ \bibnamefont
  {Jaffe}},\ }\bibfield  {title} {\enquote {\bibinfo {title} {{$\bar p p$
  Physics in the milli-TeV Region}},}\ }\bibfield  {booktitle} {\emph {\bibinfo
  {booktitle} {{1st Workshop on Antimatter Physics at Low-energy Batavia,
  Illinois, April 10-12, 1986}}},\ }\href@noop {} {\bibfield  {journal}
  {\bibinfo  {journal} {eConf}\ }\textbf {\bibinfo {volume} {C860410}},\
  \bibinfo {pages} {1} (\bibinfo {year} {1986})}\BibitemShut {NoStop}%
\bibitem [{\citenamefont {{Frederico}}\ \emph {et~al.}(2006)\citenamefont
  {{Frederico}}, \citenamefont {{Yamashita}}, \citenamefont {{Delfino}},\ and\
  \citenamefont {{Tomio}}}]{2006FBS....38...57F}%
  \BibitemOpen
  \bibfield  {author} {\bibinfo {author} {\bibfnamefont {T.}~\bibnamefont
  {{Frederico}}}, \bibinfo {author} {\bibfnamefont {M.~T.}\ \bibnamefont
  {{Yamashita}}}, \bibinfo {author} {\bibfnamefont {A.}~\bibnamefont
  {{Delfino}}}, \ and\ \bibinfo {author} {\bibfnamefont {L.}~\bibnamefont
  {{Tomio}}},\ }\bibfield  {title} {\enquote {\bibinfo {title} {{Structure of
  Exotic Three-Body Systems}},}\ }\href {\doibase 10.1007/s00601-005-0127-x}
  {\bibfield  {journal} {\bibinfo  {journal} {Few-Body Systems}\ }\textbf
  {\bibinfo {volume} {38}},\ \bibinfo {pages} {57--62} (\bibinfo {year}
  {2006})},\ \Eprint {http://arxiv.org/abs/nucl-th/0511080} {nucl-th/0511080}
  \BibitemShut {NoStop}%
\bibitem [{\citenamefont {Hasenfratz}\ and\ \citenamefont
  {Kuti}(1978)}]{Hasenfratz:1977dt}%
  \BibitemOpen
  \bibfield  {author} {\bibinfo {author} {\bibfnamefont {Peter}\ \bibnamefont
  {Hasenfratz}}\ and\ \bibinfo {author} {\bibfnamefont {Julius}\ \bibnamefont
  {Kuti}},\ }\bibfield  {title} {\enquote {\bibinfo {title} {{The Quark Bag
  Model}},}\ }\href {\doibase 10.1016/0370-1573(78)90076-5} {\bibfield
  {journal} {\bibinfo  {journal} {Phys. Rept.}\ }\textbf {\bibinfo {volume}
  {40}},\ \bibinfo {pages} {75--179} (\bibinfo {year} {1978})}\BibitemShut
  {NoStop}%
\bibitem [{\citenamefont {Hasenfratz}\ \emph {et~al.}(1980)\citenamefont
  {Hasenfratz}, \citenamefont {Horgan}, \citenamefont {Kuti},\ and\
  \citenamefont {Richard}}]{Hasenfratz:1980jv}%
  \BibitemOpen
  \bibfield  {author} {\bibinfo {author} {\bibfnamefont {P.}~\bibnamefont
  {Hasenfratz}}, \bibinfo {author} {\bibfnamefont {R.~R.}\ \bibnamefont
  {Horgan}}, \bibinfo {author} {\bibfnamefont {J.}~\bibnamefont {Kuti}}, \ and\
  \bibinfo {author} {\bibfnamefont {J.~M.}\ \bibnamefont {Richard}},\
  }\bibfield  {title} {\enquote {\bibinfo {title} {{The Effects of Colored Glue
  in the QCD Motivated Bag of Heavy Quark - anti-Quark Systems}},}\ }\href
  {\doibase 10.1016/0370-2693(80)90491-8} {\bibfield  {journal} {\bibinfo
  {journal} {Phys. Lett.}\ }\textbf {\bibinfo {volume} {95B}},\ \bibinfo
  {pages} {299--305} (\bibinfo {year} {1980})}\BibitemShut {NoStop}%
\bibitem [{\citenamefont {Braaten}\ \emph {et~al.}(2014)\citenamefont
  {Braaten}, \citenamefont {Langmack},\ and\ \citenamefont
  {Smith}}]{Braaten:2014qka}%
  \BibitemOpen
  \bibfield  {author} {\bibinfo {author} {\bibfnamefont {Eric}\ \bibnamefont
  {Braaten}}, \bibinfo {author} {\bibfnamefont {Christian}\ \bibnamefont
  {Langmack}}, \ and\ \bibinfo {author} {\bibfnamefont {D.~Hudson}\
  \bibnamefont {Smith}},\ }\bibfield  {title} {\enquote {\bibinfo {title}
  {{Born-Oppenheimer Approximation for the $XYZ$ Mesons}},}\ }\href {\doibase
  10.1103/PhysRevD.90.014044} {\bibfield  {journal} {\bibinfo  {journal} {Phys.
  Rev.}\ }\textbf {\bibinfo {volume} {D90}},\ \bibinfo {pages} {014044}
  (\bibinfo {year} {2014})},\ \Eprint {http://arxiv.org/abs/1402.0438}
  {arXiv:1402.0438 [hep-ph]} \BibitemShut {NoStop}%
\bibitem [{\citenamefont {Fleck}\ and\ \citenamefont
  {Richard}(1989)}]{Fleck:1989mb}%
  \BibitemOpen
  \bibfield  {author} {\bibinfo {author} {\bibfnamefont {S.}~\bibnamefont
  {Fleck}}\ and\ \bibinfo {author} {\bibfnamefont {J.~M.}\ \bibnamefont
  {Richard}},\ }\bibfield  {title} {\enquote {\bibinfo {title} {{Baryons with
  double charm}},}\ }\href {\doibase 10.1143/PTP.82.760} {\bibfield  {journal}
  {\bibinfo  {journal} {Prog. Theor. Phys.}\ }\textbf {\bibinfo {volume}
  {82}},\ \bibinfo {pages} {760--774} (\bibinfo {year} {1989})}\BibitemShut
  {NoStop}%
\bibitem [{\citenamefont {Fedorov}(2017)}]{Fedorov:2017bcq}%
  \BibitemOpen
  \bibfield  {author} {\bibinfo {author} {\bibfnamefont {D.~V.}\ \bibnamefont
  {Fedorov}},\ }\bibfield  {title} {\enquote {\bibinfo {title} {{Analytic
  matrix elements with shifted correlated Gaussians}},}\ }\href {\doibase
  10.1007/s00601-016-1183-0} {\bibfield  {journal} {\bibinfo  {journal} {Few
  Body Syst.}\ }\textbf {\bibinfo {volume} {58}},\ \bibinfo {pages} {21}
  (\bibinfo {year} {2017})},\ \Eprint {http://arxiv.org/abs/1702.06784}
  {arXiv:1702.06784 [nucl-th]} \BibitemShut {NoStop}%
\bibitem [{\citenamefont {Eichten}\ and\ \citenamefont
  {Quigg}(2017)}]{Eichten:2017ffp}%
  \BibitemOpen
  \bibfield  {author} {\bibinfo {author} {\bibfnamefont {Estia~J.}\
  \bibnamefont {Eichten}}\ and\ \bibinfo {author} {\bibfnamefont {Chris}\
  \bibnamefont {Quigg}},\ }\bibfield  {title} {\enquote {\bibinfo {title}
  {{Heavy-quark symmetry implies stable heavy tetraquark mesons $Q_iQ_j \bar
  q_k \bar q_l$}},}\ }\href {\doibase 10.1103/PhysRevLett.119.202002}
  {\bibfield  {journal} {\bibinfo  {journal} {Phys. Rev. Lett.}\ }\textbf
  {\bibinfo {volume} {119}},\ \bibinfo {pages} {202002} (\bibinfo {year}
  {2017})},\ \Eprint {http://arxiv.org/abs/1707.09575} {arXiv:1707.09575
  [hep-ph]} \BibitemShut {NoStop}%
\bibitem [{\citenamefont {{Hall}}\ and\ \citenamefont
  {{Post}}(1967)}]{1967PPS....90..381H}%
  \BibitemOpen
  \bibfield  {author} {\bibinfo {author} {\bibfnamefont {R.~L.}\ \bibnamefont
  {{Hall}}}\ and\ \bibinfo {author} {\bibfnamefont {H.~R.}\ \bibnamefont
  {{Post}}},\ }\bibfield  {title} {\enquote {\bibinfo {title} {{Many-particle
  systems: IV. Short-range interactions}},}\ }\href {\doibase
  10.1088/0370-1328/90/2/309} {\bibfield  {journal} {\bibinfo  {journal}
  {Proceedings of the Physical Society}\ }\textbf {\bibinfo {volume} {90}},\
  \bibinfo {pages} {381--396} (\bibinfo {year} {1967})}\BibitemShut {NoStop}%
\bibitem [{\citenamefont {{Fisher}}\ and\ \citenamefont
  {{Ruelle}}(1966)}]{1966JMP.....7..260F}%
  \BibitemOpen
  \bibfield  {author} {\bibinfo {author} {\bibfnamefont {M.~E.}\ \bibnamefont
  {{Fisher}}}\ and\ \bibinfo {author} {\bibfnamefont {D.}~\bibnamefont
  {{Ruelle}}},\ }\bibfield  {title} {\enquote {\bibinfo {title} {{The Stability
  of Many-Particle Systems}},}\ }\href {\doibase 10.1063/1.1704928} {\bibfield
  {journal} {\bibinfo  {journal} {Journal of Mathematical Physics}\ }\textbf
  {\bibinfo {volume} {7}},\ \bibinfo {pages} {260--270} (\bibinfo {year}
  {1966})}\BibitemShut {NoStop}%
\bibitem [{\citenamefont {{L{\'e}vy-Leblond}}(1969)}]{1969JMP....10..806L}%
  \BibitemOpen
  \bibfield  {author} {\bibinfo {author} {\bibfnamefont {J.-M.}\ \bibnamefont
  {{L{\'e}vy-Leblond}}},\ }\bibfield  {title} {\enquote {\bibinfo {title}
  {{Nonsaturation of Gravitational Forces}},}\ }\href {\doibase
  10.1063/1.1664909} {\bibfield  {journal} {\bibinfo  {journal} {Journal of
  Mathematical Physics}\ }\textbf {\bibinfo {volume} {10}},\ \bibinfo {pages}
  {806--812} (\bibinfo {year} {1969})}\BibitemShut {NoStop}%
\bibitem [{\citenamefont {Nussinov}\ and\ \citenamefont
  {Lampert}(2002)}]{Nussinov:1999sx}%
  \BibitemOpen
  \bibfield  {author} {\bibinfo {author} {\bibfnamefont {Shmuel}\ \bibnamefont
  {Nussinov}}\ and\ \bibinfo {author} {\bibfnamefont {Melissa~A.}\ \bibnamefont
  {Lampert}},\ }\bibfield  {title} {\enquote {\bibinfo {title} {{QCD
  inequalitie}s},}\ }\href@noop {} {\bibfield  {journal} {\bibinfo  {journal}
  {Phys. Rept.}\ }\textbf {\bibinfo {volume} {362}},\ \bibinfo {pages}
  {193--301} (\bibinfo {year} {2002})},\ \Eprint
  {http://arxiv.org/abs/hep-ph/9911532} {hep-ph/9911532} \BibitemShut {NoStop}%
\bibitem [{\citenamefont {Basdevant}\ \emph
  {et~al.}(1990{\natexlab{a}})\citenamefont {Basdevant}, \citenamefont
  {Martin},\ and\ \citenamefont {Richard}}]{Basdevant:1989pt}%
  \BibitemOpen
  \bibfield  {author} {\bibinfo {author} {\bibfnamefont {J.~L.}\ \bibnamefont
  {Basdevant}}, \bibinfo {author} {\bibfnamefont {Andre}\ \bibnamefont
  {Martin}}, \ and\ \bibinfo {author} {\bibfnamefont {J.~M.}\ \bibnamefont
  {Richard}},\ }\bibfield  {title} {\enquote {\bibinfo {title}
  {{\lowercase{IMPROVED BOUNDS ON MANY BODY HAMILTONIANS, 1. SELFGRAVITATING
  BOSONS}}},}\ }\href {\doibase 10.1016/0550-3213(90)90594-4} {\bibfield
  {journal} {\bibinfo  {journal} {Nucl. Phys.}\ }\textbf {\bibinfo {volume}
  {B343}},\ \bibinfo {pages} {60--68} (\bibinfo {year}
  {1990}{\natexlab{a}})}\BibitemShut {NoStop}%
\bibitem [{\citenamefont {Basdevant}\ \emph
  {et~al.}(1990{\natexlab{b}})\citenamefont {Basdevant}, \citenamefont
  {Martin},\ and\ \citenamefont {Richard}}]{Basdevant:1989pv}%
  \BibitemOpen
  \bibfield  {author} {\bibinfo {author} {\bibfnamefont {J.~L.}\ \bibnamefont
  {Basdevant}}, \bibinfo {author} {\bibfnamefont {Andre}\ \bibnamefont
  {Martin}}, \ and\ \bibinfo {author} {\bibfnamefont {J.~M.}\ \bibnamefont
  {Richard}},\ }\bibfield  {title} {\enquote {\bibinfo {title} {{Improved
  Bounds on Many Body Hamiltonians. 2. Baryons From Mesons in the Quark
  Model}},}\ }\href {\doibase 10.1016/0550-3213(90)90595-5} {\bibfield
  {journal} {\bibinfo  {journal} {Nucl. Phys.}\ }\textbf {\bibinfo {volume}
  {B343}},\ \bibinfo {pages} {69--85} (\bibinfo {year}
  {1990}{\natexlab{b}})}\BibitemShut {NoStop}%
\bibitem [{\citenamefont {Basdevant}\ \emph {et~al.}(1993)\citenamefont
  {Basdevant}, \citenamefont {Martin}, \citenamefont {Richard},\ and\
  \citenamefont {Wu}}]{Basdevant:1992cm}%
  \BibitemOpen
  \bibfield  {author} {\bibinfo {author} {\bibfnamefont {Jean-Louis}\
  \bibnamefont {Basdevant}}, \bibinfo {author} {\bibfnamefont {Andre}\
  \bibnamefont {Martin}}, \bibinfo {author} {\bibfnamefont {Jean-Marc}\
  \bibnamefont {Richard}}, \ and\ \bibinfo {author} {\bibfnamefont {Tai~Tsun}\
  \bibnamefont {Wu}},\ }\bibfield  {title} {\enquote {\bibinfo {title}
  {{Optimized lower bounds in the three-body problem}},}\ }\href {\doibase
  10.1016/0550-3213(93)90240-P} {\bibfield  {journal} {\bibinfo  {journal}
  {Nucl. Phys.}\ }\textbf {\bibinfo {volume} {B393}},\ \bibinfo {pages}
  {111--125} (\bibinfo {year} {1993})}\BibitemShut {NoStop}%
\bibitem [{\citenamefont {{Benslama}}\ \emph {et~al.}(1998)\citenamefont
  {{Benslama}}, \citenamefont {{Metatla}}, \citenamefont {{Bachkhaznadji}},
  \citenamefont {{Zouzou}}, \citenamefont {{Krikeb}}, \citenamefont
  {{Basdevant}}, \citenamefont {{Richard}},\ and\ \citenamefont
  {{Wu}}}]{1998FBS....24...39B}%
  \BibitemOpen
  \bibfield  {author} {\bibinfo {author} {\bibfnamefont {A.}~\bibnamefont
  {{Benslama}}}, \bibinfo {author} {\bibfnamefont {A.}~\bibnamefont
  {{Metatla}}}, \bibinfo {author} {\bibfnamefont {A.}~\bibnamefont
  {{Bachkhaznadji}}}, \bibinfo {author} {\bibfnamefont {S.~R.}\ \bibnamefont
  {{Zouzou}}}, \bibinfo {author} {\bibfnamefont {A.}~\bibnamefont {{Krikeb}}},
  \bibinfo {author} {\bibfnamefont {J.-L.}\ \bibnamefont {{Basdevant}}},
  \bibinfo {author} {\bibfnamefont {J.-M.}\ \bibnamefont {{Richard}}}, \ and\
  \bibinfo {author} {\bibfnamefont {T.~T.}\ \bibnamefont {{Wu}}},\ }\bibfield
  {title} {\enquote {\bibinfo {title} {{Optimized Lower Bound for Four-Body
  Hamiltonians}},}\ }\href {\doibase 10.1007/s006010050075} {\bibfield
  {journal} {\bibinfo  {journal} {Few-Body Systems}\ }\textbf {\bibinfo
  {volume} {24}},\ \bibinfo {pages} {39--54} (\bibinfo {year}
  {1998})}\BibitemShut {NoStop}%
\bibitem [{\citenamefont {{Boudjemaa}}\ and\ \citenamefont
  {{Zouzou}}(2009)}]{2009FBS....46..199B}%
  \BibitemOpen
  \bibfield  {author} {\bibinfo {author} {\bibfnamefont {K.-E.}\ \bibnamefont
  {{Boudjemaa}}}\ and\ \bibinfo {author} {\bibfnamefont {S.~R.}\ \bibnamefont
  {{Zouzou}}},\ }\bibfield  {title} {\enquote {\bibinfo {title} {{Optimized
  Lower Bounds for Five-Body Hamiltonians}},}\ }\href {\doibase
  10.1007/s00601-009-0061-4} {\bibfield  {journal} {\bibinfo  {journal}
  {Few-Body Systems}\ }\textbf {\bibinfo {volume} {46}},\ \bibinfo {pages}
  {199--220} (\bibinfo {year} {2009})}\BibitemShut {NoStop}%
\bibitem [{\citenamefont {Semay}\ and\ \citenamefont
  {Silvestre-Brac}(1994)}]{Semay:1994ht}%
  \BibitemOpen
  \bibfield  {author} {\bibinfo {author} {\bibfnamefont {C.}~\bibnamefont
  {Semay}}\ and\ \bibinfo {author} {\bibfnamefont {B.}~\bibnamefont
  {Silvestre-Brac}},\ }\bibfield  {title} {\enquote {\bibinfo {title}
  {{Diquonia and potential models}},}\ }\href {\doibase 10.1007/BF01413104}
  {\bibfield  {journal} {\bibinfo  {journal} {Z. Phys.}\ }\textbf {\bibinfo
  {volume} {C61}},\ \bibinfo {pages} {271--275} (\bibinfo {year}
  {1994})}\BibitemShut {NoStop}%
\bibitem [{\citenamefont {Bhaduri}\ \emph {et~al.}(1981)\citenamefont
  {Bhaduri}, \citenamefont {Cohler},\ and\ \citenamefont
  {Nogami}}]{Bhaduri:1981pn}%
  \BibitemOpen
  \bibfield  {author} {\bibinfo {author} {\bibfnamefont {R.~K.}\ \bibnamefont
  {Bhaduri}}, \bibinfo {author} {\bibfnamefont {L.~E.}\ \bibnamefont {Cohler}},
  \ and\ \bibinfo {author} {\bibfnamefont {Y.}~\bibnamefont {Nogami}},\
  }\bibfield  {title} {\enquote {\bibinfo {title} {{A Unified Potential for
  Mesons and Baryons}},}\ }\href {\doibase 10.1007/BF02827441} {\bibfield
  {journal} {\bibinfo  {journal} {Nuovo Cim.}\ }\textbf {\bibinfo {volume}
  {A65}},\ \bibinfo {pages} {376--390} (\bibinfo {year} {1981})}\BibitemShut
  {NoStop}%
\bibitem [{\citenamefont {Kiselev}\ \emph {et~al.}(2017)\citenamefont
  {Kiselev}, \citenamefont {Berezhnoy},\ and\ \citenamefont
  {Likhoded}}]{Kiselev:2017eic}%
  \BibitemOpen
  \bibfield  {author} {\bibinfo {author} {\bibfnamefont {V.~V.}\ \bibnamefont
  {Kiselev}}, \bibinfo {author} {\bibfnamefont {A.~V.}\ \bibnamefont
  {Berezhnoy}}, \ and\ \bibinfo {author} {\bibfnamefont {A.~K.}\ \bibnamefont
  {Likhoded}},\ }\bibfield  {title} {\enquote {\bibinfo {title} {{Quark-diquark
  structure and masses of doubly charmed baryons}},}\ }\href@noop {} {\
  (\bibinfo {year} {2017})},\ \bibinfo {note} {to appear in Phys. Atomic
  Nuclei},\ \Eprint {http://arxiv.org/abs/1706.09181} {arXiv:1706.09181
  [hep-ph]} \BibitemShut {NoStop}%
\end{thebibliography}
%
\end{document}